\documentclass[aps,prl,superscriptaddress,twocolumn,10pt,longbibliography]{revtex4-1}

\usepackage{graphicx}
\usepackage{amsmath}
\usepackage{amssymb}
\usepackage{dsfont}
\usepackage{xspace}
\usepackage{color}
\usepackage[usenames,dvipsnames]{xcolor}
\usepackage[colorlinks=true,linkcolor=blue,urlcolor=blue,citecolor=blue]{hyperref}
\usepackage{bbm}

\usepackage{stmaryrd}
\usepackage[normalem]{ulem}
\usepackage{comment}

\newcommand{\eref}[1]{(\ref{#1})}
\newcommand{\fref}[1]{Fig.\,\ref{#1}}

\newcommand{\ket}[1]{ \left | #1\right \rangle}

\newcommand{\braket}[2]{ \left \langle #1\middle| #2\right \rangle}

\newcommand{\dd}{ {\rm d} }

\newcommand{\ii}{ {\rm i} }

\newcommand{\ave}[1]{{\langle #1\rangle}}

\def\sz{\sigma^{\rm z}}
\def\sx{\sigma^{\rm x}}

\newcommand{\eqw}[1]{(\ref{#1})}
\newcommand{\eq}[1]{Eq.~(\ref{#1})}

\newcommand{\fig}[1]{Fig.\thinspace{}\ref{#1}}

\newcommand{\fc}[1]{({#1})}
\newcommand{\figc}[2]{Fig.\thinspace{}\ref{#1}\thinspace{}\fc{#2}}

\begin{document}

\title{Dynamical Quantum Phase Transitions in Spin Chains with Long-Range Interactions: Merging different concepts of non-equilibrium criticality}

\author{Bojan \v{Z}unkovi\v{c}}
\address{SISSA --- International School for Advanced Studies, via Bonomea 265, 34136 Trieste, Italy}
\author{Markus Heyl}
\affiliation{Department of Physics, Walter Schottky Institute, and Institute for Advanced Study, Technical University of Munich, 85748 Garching, Germany}%
\affiliation{Max-Planck-Institut f\"ur Physik komplexer Systeme, 01187 Dresden, Germany}%
\author{Michael Knap} 
\affiliation{Department of Physics, Walter Schottky Institute, and Institute for Advanced Study, Technical University of Munich, 85748 Garching, Germany}%
\author{Alessandro Silva}
\address{SISSA --- International School for Advanced Studies, via Bonomea 265, 34136 Trieste, Italy}

\begin{abstract}
We theoretically study the dynamics of a transverse-field Ising chain with power-law decaying interactions characterized by an exponent $\alpha$, which can be experimentally realized in ion traps. We focus on two classes of emergent dynamical critical phenomena following a quantum quench from a ferromagnetic initial state: The first one manifests in the time averaged order parameter, which vanishes at a critical transverse field. We argue that such a transition occurs only for long-range interactions $\alpha \leq 2$ . The second class corresponds to the emergence of time-periodic singularities in the return probability to the ground state manifold (a.k.a. Loschmidt echo) which is obtained for all values of $\alpha$ and agrees with the order parameter transition for $\alpha\leq 2$. We characterize how the two classes of nonequilibrium criticality correspond to each other and give a physical interpretation based on the symmetry of the time-evolved quantum states.
\end{abstract}

\date{\today}

\maketitle

Recent experiments with cold atoms~\cite{hofferberth_07, Gring2012, langen2013, Hild2014, Schreiber2015oo, Bordia2016, Choi2016, Bordia2016pd} and trapped ions~\cite{Smith2016, Neyenhuis2016a, Martinez2016} have realized nonequilibrium quantum states with exotic properties that cannot be captured by a thermodynamic equilibrium description. This includes the observation of prethermalization~\cite{hofferberth_07, Gring2012, langen2013, Neyenhuis2016a} and many-body localization~\cite{Schreiber2015oo, Bordia2016, Bordia2016pd, Smith2016, Choi2016}. Despite these remarkable discoveries, it is still a major challenge to reveal universal properties of nonequilibrium quantum states. One possible approach for developing a general understanding of far-from-equilibrium dynamics is to explore concepts of nonequilibrium critical phenomena. However, due to the lack of clear generic principles, different concepts of dynamical criticality have been introduced~\cite{Yuzbashyan2006, Diehl2008, barmettler, Eckstein2009wj, Diehl2010, SB10, Garrahan2010xw, Mitra2012, Heyl2012b}.

\begin{figure}
\includegraphics[width=0.48\textwidth]{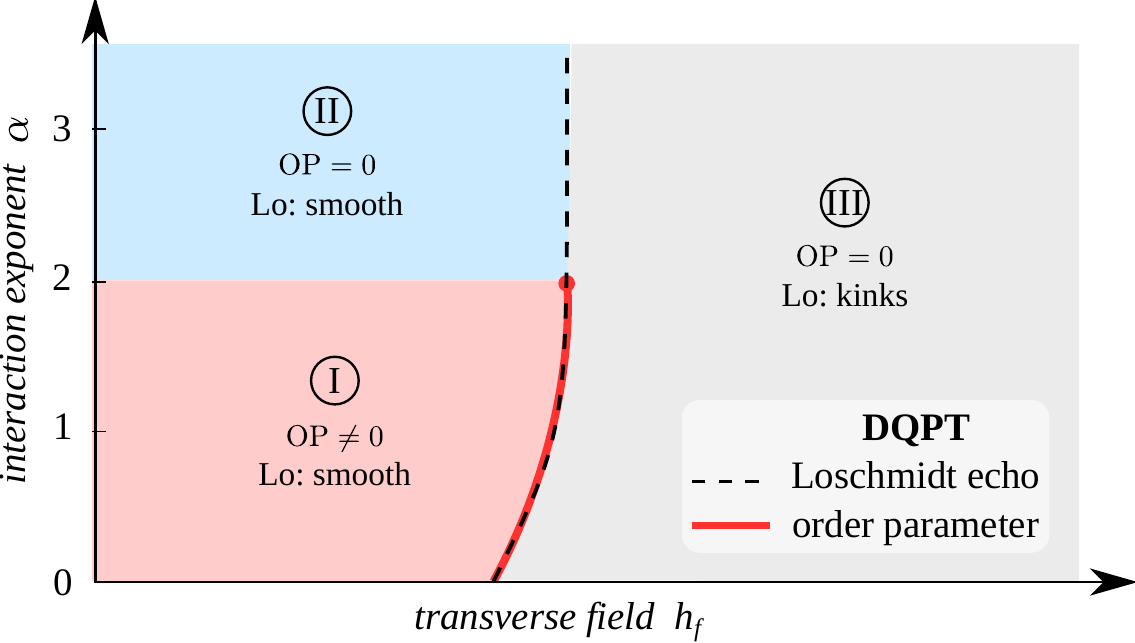}
\caption{\textbf{Dynamical phase diagram.} We study the quantum dynamics of an Ising chain with power-law decaying interactions by preparing the system in a fully polarized state and abruptly switching on a finite transverse field $h_f$. We identify the dynamical quantum phase transition (DQPT) through two mechanisms: One introduces an order parameter (DQPT-OP) that is finite only in the dynamical ferromagnetic phase, whereas the other is based on non-analytic kinks in the Loschmidt rate function (DQPT-LO) that only arise for quenches across the transition. When the interaction exponent $\alpha<2$, the DQPT occurs simultaneously for both cases along the red line, bridging the dynamical ferromagnetic (I) and the dynamical paramagnetic (III) phase. We argue that for $\alpha>2$ the system does not establish a finite order parameter and thus the DQPT-OP ends at $\alpha=2$. Yet, the DQPT-LO persists for arbitrarily large $\alpha$ (dashed line) separating two dynamical phases characterized by a monotonic decay (II) and an oscillating decay (III) of the magnetization.}
\label{Fig1}
\end{figure}

In this work, we show that two seemingly unrelated nonequilibrium critical phenomena are actually intimately connected. In particular, the first class of nonequilibrium criticality describes dynamical quantum phase transitions (DQPT) in the asymptotic late-time steady state of an order parameter (DQPT-OP) that is finite in one dynamical phase but vanishes in the other~\cite{Yuzbashyan2006, SB10}. The second class, are DQPTs associated with singular behavior in the transient real-time evolution of Loschmidt echoes (DQPT-LO)~\cite{Heyl2012b, Heyl2014}. By studying the quantum dynamics of an initially fully polarized state in a transverse-field Ising chain with power-law decaying interactions, we show that these two types of nonequilibrium critical phenomena are related in several ways: First, they predict consistent values for the dynamical critical point, see \fig{Fig1}. Second, the singularities in the Loschmidt echo are related to zeros in the time evolution of the order parameter. Third, we argue that the dynamics restores the symmetry breaking imprinted by the initial polarized state only when crossing the DQPT-LO but ceases to do so for quenches within the same dynamical phase.

\textit{\textbf{Model and Protocol.---}}Experimentally, the real-time dynamics of long-range interacting spin chains in a transverse field can be studied with trapped ions~\cite{Lanyon2011, Britton2012, Jurcevic2014, Richerme2014a, Smith2016, Martinez2016} where power-law decaying interactions between the effective spins are mediated by collective vibrations of the underlying ionic crystal ($0\leq \alpha \leq 3$~\cite{Islam2013}). The corresponding Hamiltonian is
\begin{eqnarray}
\hat H(h)=-\sum_{i\not= j=1}^N V(i-j)\sx_i\sx_j-h \sum_{j=1}^N\sz_j,
\label{eq:h}
\end{eqnarray}
with the transverse field $h$ and the interaction potential $V(x)=J v(x)/N(\alpha)$. Here, $v(x)= |x|^{-\alpha}$ describes the power-law decaying interactions and $J$ sets the interaction strength. We added a normalization constant
$N(\alpha)=\frac{1}{N-1}\sum_{i\not= j=1}^N v(i-j)$  that ensures the intensive scaling of the energy density for any $\alpha$. For all values of $\alpha$, this model is known to display an equilibrium quantum phase transition from a ferromagnet to a paramagnet. At finite temperatures, the equilibrium ferromagnetic phase is in one dimension only stable for $\alpha \leq 2$~\cite{dutta_phase_2001,knap_probing_2013}.

\begin{figure}
\includegraphics[width=0.48\textwidth]{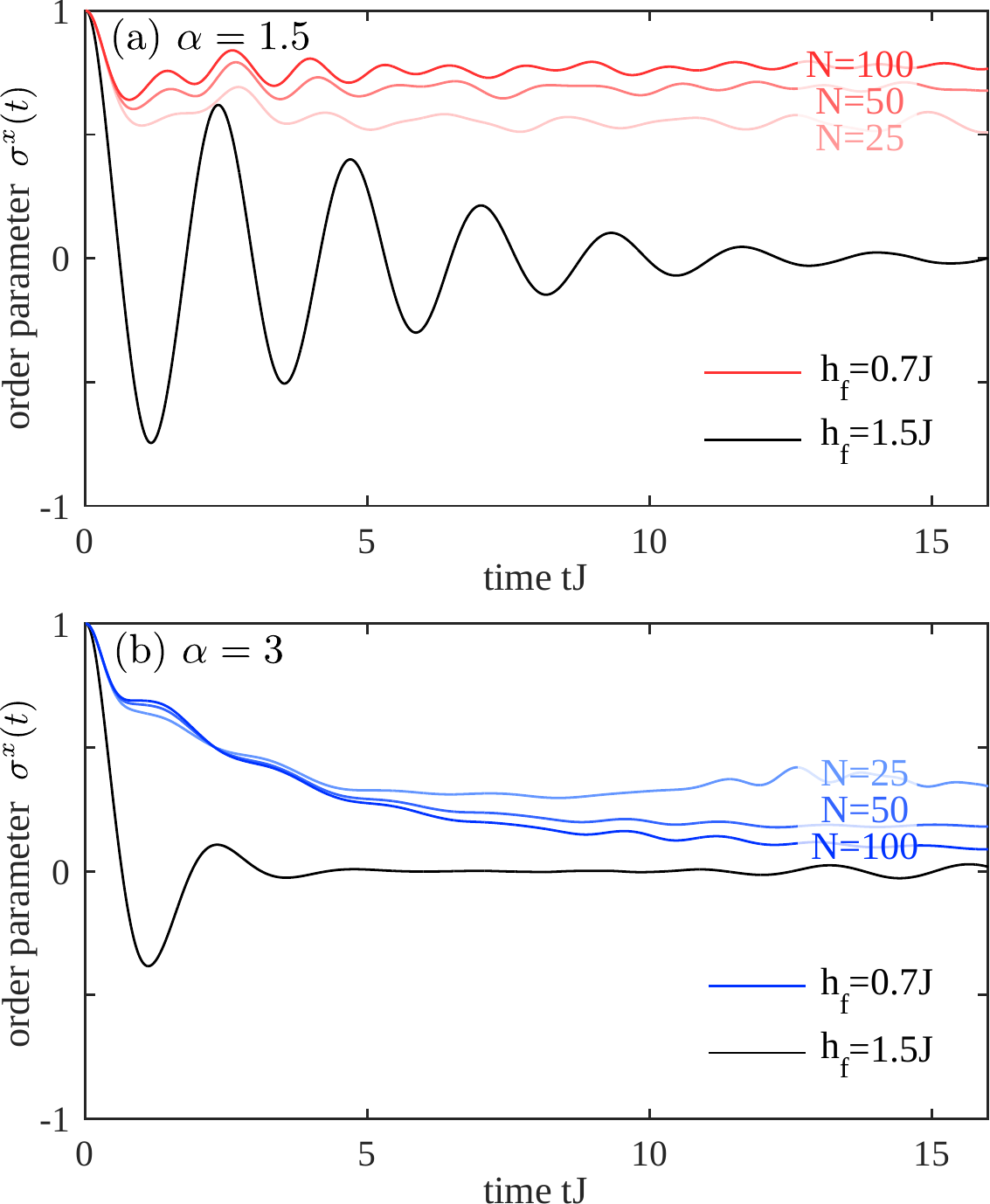}
\caption{\textbf{Time evolving the order parameter.} Starting from an initially fully polarized state in $x$-direction we quench the system to a finite transverse field $h_f$ and calculate the ensuing dynamics of the order parameter $\sigma^x(t)$. \fc{a} For an interaction exponent $\alpha=1.5$ and weak transverse field $h_f=0.7J$ we find that the order parameter assumes a finite value at late times and hence describes a dynamical symmetry-broken state with ferromagnetic order, whereas it decays to zero with strong oscillations for quenches to large transverse field $h_f=1.5J$. \fc{b} Even though for shorter-ranged interactions $\alpha=3$ the order parameter reaches zero at late times for all values of $h_f$, the nature of the decay is very different: For small fields $h_f=0.7J$ it decays with a time scale much longer than the microscopic scales, whereas for large fields $h_f=1.5J$ it oscillates around zero and decays rapidly. }
\label{fig2}
\end{figure} 
\begin{figure*}
\includegraphics[width=0.98\textwidth]{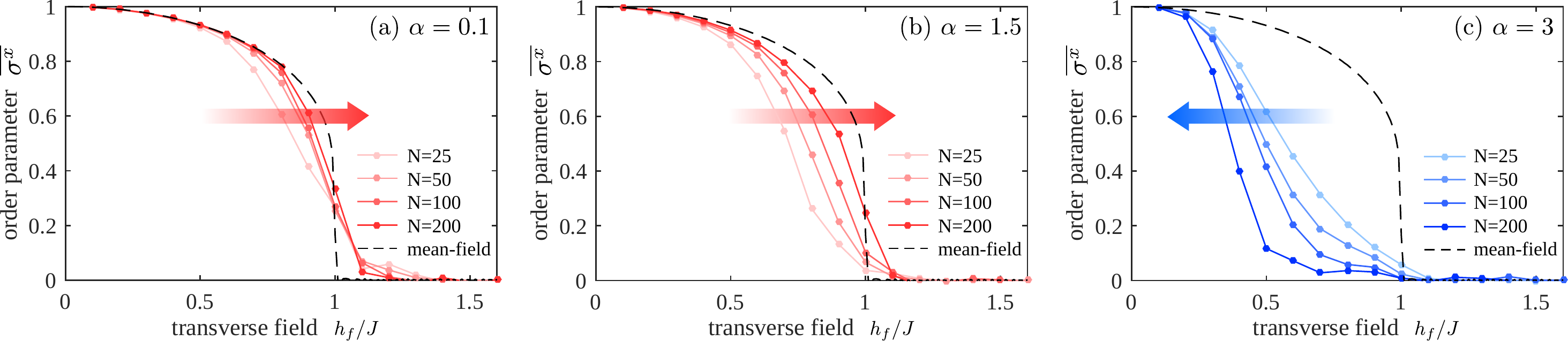}
\caption{\textbf{Dynamical phase diagram of the order parameter.} We estimate the asymptotic value of the order parameter $\overline{\sigma^x}$ as a function of the quenched transverse field $h_f$ for different system sizes $N$ and interaction exponents \fc{a} $\alpha=0.1$, \fc{b} $\alpha=1.5$, and \fc{c} $\alpha=3$. For both values of $\alpha<2$ we find that the finite-size flow of the order parameter indicates a DQPT with the critical point $h_f \sim J$. For very long-ranged interactions, \fc{a}, the asymptotic order parameter $\overline{\sigma^x}$, approaches the mean-field predictions, $\alpha=0$, (dashed lines) with increasing system size. By contrast, for relatively short ranged interactions \fc{c} $\alpha=3$, the finite-size scaling clearly suggests that the oder parameter $\overline{\sigma^x}$ flows toward zero in the thermodynamic limit for all values of the transverse field. }
\label{fig3}
\end{figure*}

We are studying the quantum dynamics following a global quantum quench in the transverse field $h$. Initially, we prepare the system in the fully polarized state $|+\rangle = |\rightarrow \dots \rightarrow\rangle$ which is one of the two ground states of Hamiltonian \eqw{eq:h} for initial field $h_i=0$. Such a state can be realized with high fidelity in systems of trapped ions~\cite{Lanyon2011}. We then suddenly switch on a transverse field $h_f>0$ and monitor the ensuing real-time dynamics  governed by the Hamiltonian $\hat H(h_f)$. The time evolution is computed  numerically using a recently developed algorithm~\cite{Haegeman2011} based on a time-dependent variational principle. All presented data is evaluated for matrix product state bond dimension $100$ and time step $0.02/J$; see Supplementary Material for details on the algorithm and convergence~\cite{supp}.

\textbf{\textit{Time evolution of the order parameter.---}}The first class of dynamical criticality, DQPT-OP, occurs in the long-time asymptotics of a dynamical order parameter~\cite{Yuzbashyan2006, Eckstein2010, Gambassi2010, SB10, SB11, SB13, Smacchia2015}, which is finite for quenches within the ordered phase $h_f<h_c$ and zero for quenches across the dynamical transition $h_f>h_c$. For our model the order parameter is the time-averaged longitudinal magnetization
\begin{eqnarray}
\overline{\sx}=\lim_{T\rightarrow+\infty}\frac{1}{T}\int_0^Tdt \, \sigma^x(t),
\end{eqnarray}
with $\sigma^\beta(t) = \langle S^\beta(t) \rangle$ ($\beta=x,y,z$) and  $S^{\beta}=1/N\sum_{i}\sigma^{\beta}_i$ denoting the collective spin operators. The order parameter coincides with the asymptotic long-time limit of the magnetization whenever it relaxes to a constant value. 

DQPT-OPs have been studied extensively in various integrable quantum many-body systems such as BCS models~\cite{Yuzbashyan2006}, models with infinite-range interactions~\cite{Eckstein2010,Hamerla2013,Schiro2010,Gambassi2010,SB10,SB11,SB13} as well as field theories in high dimensions~\cite{SB13,Smacchia2015}. Due to dynamical constraints imposed by integrability, these systems are not thermalizing and  the asymptotic long-time steady state is described by a generalized Gibbs ensemble. As a consequence these DQPT-OPs do not reduce to thermal transitions and exhibit properties not accessible in equilibrium states.

A simple, analytically tractable regime of our model is the infinite-range limit, $\alpha=0$. There, the dynamics of the order parameter corresponds to the precession of a single collective spin $S^{\beta}$, implying that $\sx(t)$ oscillates persistently in time with a single frequency around a mean value set by $\overline{\sx}$. Initializing the system in the ferromagnetic ground state, $h_i<h_c$, a DQPT-OP can occur, characterized by an order parameter $\overline{\sx}$ that remains finite for $h_f<h_c$ but is zero for $h_f>h_c$. For $\alpha=0$, the critical value of the dynamical transition can be computed analytically $h_c=J+h_i/2$~\cite{SB11,Zunkovic2016a}. 

When increasing the interaction range, the order parameter still remains finite for quenches within the dynamical ferromagnetic phase, see \figc{fig2}{a} for $\alpha=1.5$. When crossing the DQPT-OP, the persistent oscillations observed for the infinite-range model get damped and decay to zero. Increasing $\alpha$ even further, we observe a decay of the order parameter regardless of the final transverse field $h_f$, see \figc{fig2}{b} for $\alpha=3$. This is consistent with the expectation that a spin chain with sufficiently short-ranged interactions cannot feature an ordered non-equilibrium steady-state.

\textbf{\textit{Dynamical transition in the order parameter.---}}We now focus on the steady-state value of the order parameter $\overline{\sigma^x}$, which in finite size systems always shows the same qualitative crossover:  a monotonic decrease of $\overline{\sx}$ from one to zero as the final transverse field $h_f$ is increased. However, upon finite-size scaling we observe a markedly different behavior when tuning the value of $\alpha$ (\fig{fig3}). For small $\alpha$ and moderate fields $h_f \lesssim J$, \figc{fig3}{a,b}, $\overline{\sx}$ increases with system size $N$ and the finite-size flow suggests that the critical point is close to $h_c \approx J$ in the thermodynamic limit. By contrast, for large $\alpha$, \figc{fig3}{c}, the order parameter $\overline{\sx}$ rapidly vanishes with increasing system size. Therefore, a DQPT-OP cannot exist when the interactions are sufficiently short-ranged. We argue for $\alpha_c=2$ being the upper bound for observing DQPT-OP (see Supplemental Material~\cite{supp} for additional data) even though there a exists a crossover region between $2 \leq \alpha \leq 2.4$ in which the finite-size flow is not fully indicative. 

Although one might expect that integrability is broken for interactions with any finite range $\alpha>0$, we show that the DQPT-OP is not a thermal transition in the limit of small but finite $\alpha$. To this end, we first compute the critical field $h_c^\text{th}$ at which the energy deposited by the quantum quench corresponds to the energy of the thermal equilibrium transition. For $\alpha=0$ we find $h_c^\text{th} = \sqrt{2} J$~\cite{supp}, which we assume to change only perturbatively for slightly larger $\alpha$. Comparing this with the dynamical critical field $h_c \approx J$, we can conclude that $h_c$ and $h_c^\text{th}$ are incompatible and thus the transition at $\alpha=0.1$ is not thermal. Determining the value of $\alpha$ at which the system starts to thermalize remains an exciting open question.

\textbf{\textit{Dynamical transition in the Loschmidt echo.---}}The second class of dynamical transitions we consider are DQPT-LOs, that arise as singularities in Loschmidt amplitudes $\mathcal{G}(t)= \langle \Psi |\exp[-iH(h_f)t] | \Psi \rangle$ as a function of time~\cite{Heyl2012b}, where $|\Psi\rangle$ denotes the initial state. Formally, Loschmidt amplitudes at imaginary times resemble equilibrium boundary partition functions~\cite{Gambassi2011,GS12,Heyl2012b}. Therefore, it is suitable to introduce a dynamical counterpart of the free energy density, which is the Loschmidt rate function (or large deviation function~\cite{GS12}) $g(t) = - N^{-1} \log[\mathcal{G}(t)]$. Similarly to equilibrium free energies being nonanalytic at conventional phase transitions, also the Loschmidt rate function $g(t)$ can display non-analyticities, which defines the DQPT-LO. Equivalently, DQPT-LOs occur in the Loschmidt echo $\mathcal{L}(t)=|\mathcal{G}(t)|^2$ which is the probability associated with the amplitude $\mathcal{G}(t)$. Such DQPT-LOs have been observed in different models~\cite{pollmann_10, Heyl2012b, Karrasch2013, Andraschko2014, Kriel2014, Canovi2014, Schmitt2015, Heyl2014, Vajna2014, Vajna2014tc, Abeling2016, Budich2016, Sharma2016, Huang2016, Flaschner2016, Heyl2015,Heyl2016} and have recently been measured experimentally~\cite{Flaschner2016}.

\begin{figure}
\includegraphics[width=0.48\textwidth]{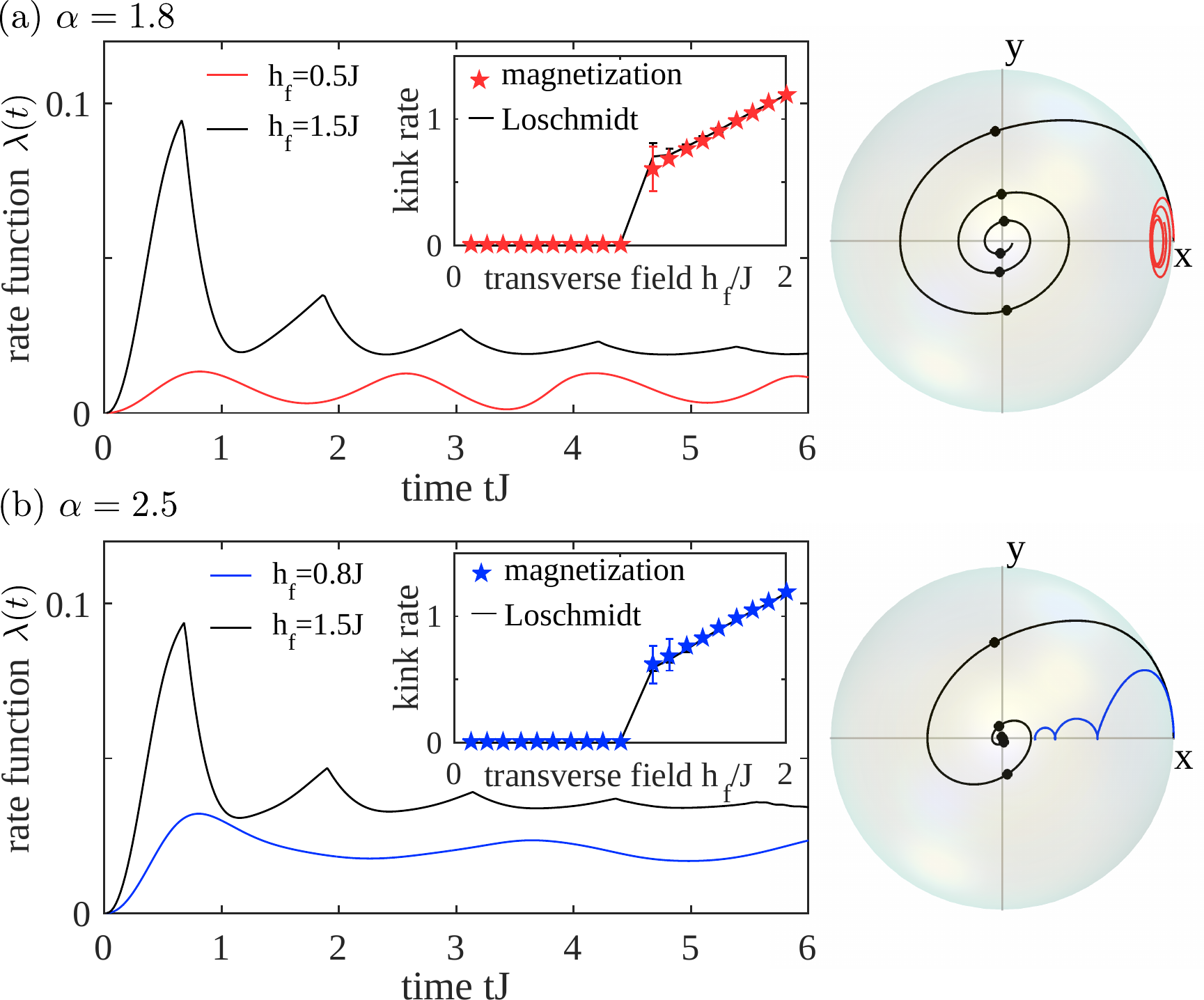}
\caption{\textbf{Dynamical quantum phase transitions in the Loschmidt echo.} We compute an extension of the Loschmidt echo, which is the return probability to the degenerate ground state manifold, \eq{timeop}, for different values of the transverse field $h_f$ and interaction exponent $\alpha$. We observe non-analyticities in the associated rate function $\lambda(t)$, for arbitrary values of the interaction exponent $\alpha$, provided the final transverse field $h_f$ is sufficiently large.  The insets show the typical rate of kinks in $\lambda(t)$, solid line, and compares it to the zero crossings of the order parameter $\sigma^x(t)$, which agree for both values of $\alpha$ within errorbars. The right panels show the evolution of the magnetization $\vec \sigma(t)$ projected onto the $xy$-plane of the Bloch sphere. In order to illustrate the connection between DQPT-OPs and DQPT-LOs we include as dots the points where kinks occur. When quenching across the DQPT-LO, the magnetization spreads over both hemispheres (black curves) whereas it remains located on one hemisphere for quenches within the same dynamical phase (blue and red curves), indicating a bifurcation of the dynamics.}
\label{fig4}
\end{figure}

The Loschmidt amplitude is not uniquely defined when the ground-state manifold of the initial Hamiltonian is degenerate. 
In order to maintain the connection of DQPT-LOs to macroscopic observables and therefore potentially to DQPT-OPs, the proper generalization is the probability to stay in the ground state manifold~\cite{Heyl2014}
\begin{eqnarray}\label{timeop}
P(t)=\sum_{n}|\langle \Psi_n(h_i) |e^{-iH(h_f)t} | \Psi_0(h_i) \rangle|^2,
\end{eqnarray}
which reduces to the Loschmidt echo $\mathcal{L}(t)$ in the limit of a single ground state. 
Here, $\{|\Psi_n(h_i)\rangle\}$ denotes the degenerate states at $h_i$ and $| \Psi_0(h_i) \rangle$ is the chosen initial condition. In our case we have that $| \Psi_0(h_i) \rangle = |+\rangle $ and we denote $| \Psi_1(h_i) \rangle = |-\rangle = |\leftarrow\ldots\leftarrow\rangle$. Consequently, we get $P(t) = P_+(t) + P_-(t)$ with $P_+(t) = |\langle + | +(t) \rangle|^2$ and $P_-(t) = |\langle - | +(t) \rangle|^2$.

\textbf{\textit{Merging the different concepts of DQPT.---}}Let us now establish the connection of the two concepts of dynamical criticality. For this purpose we first consider the limit of $\alpha=0$ where the dynamics is described by semi-classical Bloch equations for the collective spin $\vec{\sigma}(t) = \{\sigma^x(t),\sigma^y(t),\sigma^z(t)\}$. In that case the individual probabilities $P_\pm(t) = \exp[-N \lambda_\pm(t)]$ with $\lambda_\pm(t)=-\log[(1\pm \vec{\sigma}(t)\cdot \vec{\sigma}(0))/2]$~\cite{Bojan2016} exhibit a particularly illustrative form. In particular, for the fully polarized state, $\vec{\sigma}(t)\cdot \vec{\sigma}(0)$ measures the projection of $\vec{\sigma}(t)$ onto the $x$-axis.

DQPT-LOs can occur in $P(t)$ because the individual probabilities $P_\pm(t)=\exp[-N\lambda_\pm(t)]$ show an exponential dependence on system size $N$. Therefore, in the thermodynamic limit only one of the two $P_\pm(t)$ dominates such that $P(t) = \exp[-N\lambda(t)]$ with $\lambda(t) = \min_{\eta=\pm} \lambda_\eta(t)$~\cite{Heyl2014}. While at short times $\lambda(t) = \lambda_+(t)$ due to the initial condition,  $\lambda_-(t)$ can take over at a critical time $t_c$, which leads to a kink in $\lambda(t)$. For the concrete case of  $\alpha=0$ this can be traced back to $\vec{\sigma}(t)$ crossing the equator of the Bloch sphere, $\sigma^x=0$, because then $\lambda_+ = \lambda_-$. As we have seen before (Fig.~\ref{fig2}), this can happen only when the DQPT-OP is crossed, i.e., for $h_f>h_c$. Therefore, a DQPT-LO  occurs only when crossing the DQPT-OP, which manifests itself in a vanishing long-time magnetization $\overline{\sigma^x}=0$. In this way the $\mathbb{Z}_2$ symmetry, broken explicitly by the initial state, is restored in the long-time limit as well as at the critical times at which the DQPT-LO occur. The restoration of this $\mathbb{Z}_2$ symmetry is accompanied by a change in the symmetry of the trajectory of the magnetization on the Bloch sphere; right panels in Fig.~\ref{fig4}.

Although, these considerations address a fine-tuned limit of $\alpha=0$, we show in Fig.~\ref{fig4} based on numerical data that the relation between DQPT-LO and DQPT-OP extends also to $\alpha>0$. DQPT-LOs in the form of kinks occur whenever the system is quenched sufficiently strongly such that $h_f>h_c$ whereas for $h_f<h_c$ the rate function $\lambda(t)$ stays smooth. In the insets of \fig{fig4} we compare the time scale for the periodicity of the DQPT-LOs with the period of the zeros of $\sigma_x$. Specifically, we plot the inverse of these time scales and find within the numerical error bars very good agreement over a wide range of $h_f$ suggesting a very close connection of $\sigma^x(t)$ and DQPT-LOs irrespective of $\alpha$. The precise location of the zeros in $\sigma^x$ exhibits an essentially constant shift compared to the kinks in $\lambda(t)$~\cite{Heyl2012b,Heyl2014}. This is illustrated in the Bloch spheres of Fig. 4, where the kinks (black dots) appear slightly later in time than the zero crossings of the order parameter $\sigma^x=0$. Moreover, we emphasize that the connection between DQPT-LOs and the zeros of $\sigma^x$ is also valid for $\alpha>2$ where no DQPT-OP occurs. The field $h_c$ marking the appearance of DQPT-LOs for $h_f>h_c$ then separates a regime of monotonic decay of $\sigma^x$ for $h_f<h_c$ from oscillatory decay for $h_f>h_c$;  Fig.~\ref{fig2}.

\textbf{\textit{Conclusions and Outlook.---}}We have studied dynamical quantum phase transition in a transverse field Ising chain with power-law decaying interactions. We have argued that two seemingly different concepts of non-equilibrium criticality, specifically dynamical transitions in the late-time behavior of the order parameter and dynamical transitions in the Loschmidt echo, are actually intimately related in the following ways: (i) We find that both of them predict consistent values for the dynamical critical point for interaction exponent $\alpha<2$. (ii) For generic values of $\alpha$, the period of kinks in the Loschmidt rate function agrees with the period of zeros in the order parameter. (iii) The order parameter restores symmetry imprinted by the initial polarized state, only for quenches across the dynamical quantum phase transition but ceases to do so for quenches within the same dynamical phase. 

In future studies, it would be exciting to extract the dynamical critical exponents of the order parameter. This could be achieved by infinite system size variants of our numerical technique. Furthermore, studying in detail the scaling of order parameter fluctuations with system size could establish for which values of the interaction exponent, our system is thermalizing according to the eigenstate thermalization hypothesis.

\textit{\textbf{Acknowledgments.---}} 
We acknowledge support from the Technical University of Munich - Institute for Advanced Study, funded by the German Excellence Initiative and the European Union FP7 under grant agreement 291763, the Deutsche Akademie der Naturforscher Leopoldina under
grant number LPDR 2015-01 and by the Deutsche Forschungsgemeinschaft via the Gottfried Wilhelm Leibniz Prize programme. BZ was supported by the ERC under Starting Grant 279391 EDEQS.

\newpage
\setcounter{equation}{0}
\setcounter{figure}{0}
\renewcommand\theequation{S--\arabic{equation}}
\renewcommand\thefigure{S--\arabic{figure}}

\onecolumngrid
\setlength{\abovecaptionskip}{6pt plus 3pt minus 2pt}
\setlength{\belowcaptionskip}{0pt}
\vspace{15mm}

\begin{center}
\textbf{\large Supplementary Materials}
\end{center}

\tableofcontents
\appendix
\section{Infinite-range transverse-field Ising model}
In this section we summarize the behaviour of the infinite-range transverse-field Ising model described by the Hamiltonian
\begin{align}
\label{eq:mfH}
H=-\frac{J}{N}\sum_{j,k=1}^N\sx_j\sx_k - h\sum_{j}^N\sz_j,
\end{align}
where $\sigma_i^{\alpha}$ are Pauli matrices satisfying the communation relations $[\sigma_j^{\alpha},\sigma_j^{\beta}]=2i\epsilon^{\alpha\beta\gamma}\sigma_j^\gamma$.
After recapitulating the equilibrium properties in the mean-field picture we describe two notions of dynamical phase transitions and their relation to the bifurcation of semiclassical trajectories. We end by comparing the dynamical transition to the finite temperature transition.

\subsection{Equilibrium phase diagram}
The Hamiltonian Eq.\eref{eq:mfH} can  be conveniently rewritten as
\begin{align}
\label{eq:hbigspin}
H/N=-J(S^{\rm x})^2-h S^{\rm z},
\end{align}
with $S^a=\frac{1}{N}\sum_j\sigma^a_j,$ $(a=\rm{x,y,z})$, and $[S^{\rm x},S^{\rm y}]=\frac{2\ii}{N} S^{\rm z}$. At zero temperature the phase diagram is determined by the minima of the semi-classical energy function
\begin{align}
H(\theta,\varphi)= -J\cos^2(\theta)\cos^2(\varphi)-h\sin(\theta),
\label{eq:sch}
\end{align}
where $(S^{\rm x}, S^{\rm y}, S^{\rm z})\rightarrow(\cos\theta\cos\varphi,\cos\theta\sin\varphi,\sin\theta)$. The minima of the energy \eref{eq:sch} as a function of $\theta$ lie at $\theta = \arcsin(h/2J)$ (ferromagnetic phase) for $h/J<2$ and at $\theta=\pi/2$ for $h/J>2$ (paramagnetic phase). At finite temperature we resort to the mean-field analysis. The mean-field Hamiltonian is
\begin{align}
H_{\rm mf} =- \begin{pmatrix}
h&2J\sx\\
2J\sx&-h
\end{pmatrix},
\end{align}
where $\sigma^a=\ave{S^a}$ for  $a=\rm{x,y,z}$. At finite temperature this leads to the self-consistency equation
\begin{align}
\sx=\frac{2J \sx}{\sqrt{h^2+4J^2(\sx)^2}}\tanh\left( \frac{1}{T}\sqrt{h^2+4J^2(\sx)^2}\right),
\label{eq:tmf}
\end{align}
Equation \eref{eq:tmf} indentifies two regions in the $(T/J,h/J)$ plane, a ferromagnetic one with $\sx\neq0$ and a paramagnetic one with $\sx=0$. The boundary is given by the equation 
\begin{align}
\label{eq:Tc}
\frac{h_c^\text{th}}{2J}=\tanh(\frac{h_c^\text{th}}{T}).
\end{align}

\subsection{Dynamical phase transition}
In this section we discuss the semiclassical picture of the dynamical transitions and compare it to the exact finite size scaling of the infinite-range transverse-field Ising model.

\subsubsection{Semiclassical analysis}
In the large $N$ limit the mean-field picture provides the correct dynamical behaviour of local observables. Starting from a semiclassical state (for example a ground state at some $h_i$), where all spins are pointing in one direction determined by two angles $\theta_i$ and $\varphi_i$, all spins evolve according to the semiclassical equations and therefore keep pointing in the same direction. In this situation the state can be described as a single big spin, the dynamics of which can conveniently be visualised on the Bloch sphere (see \fref{fig3}). 

Let us first consider quenches starting in the ferromagnetic phase; $h_i<2$. We observe two types of trajectories, which can be characterised by the time-averaged magnetisation $\overline{\sx}$ \cite{SB11,Zunkovic2016a}. Trajectories of the first type occur for small enough $h_f$ and are restricted to a single hemisphere giving rise to a non-vanishing magnetisation $\sx$. Trajectories of the second type in turn occur for large $h_f$ and are symmetric under the transformation $x\rightarrow-x$ with a vanishing time-averaged magnetisation $\overline{\sx}$. The critical magnetic field separating these two regions can be calculated from the condition $H_f(\theta_i,\varphi_i)=H_f(\pi/2,0)$ and is given by $h_c=J+h_i/2$.

The transition between the symmetric and symmetry-broken trajectories can also be captured by another notion of dynamical criticality \cite{Heyl2012b,Heyl2014} consisting in the observation of singularities in time of the probability to return to the ground state manifold (in short return probability). To demonstrate singular behaviour of the return probability in the infinite-range Ising model we consider first the quench from $h_i=0$. The overlap between two semiclassical states is given by the formula
\begin{align}
\label{eq:overlap}
\braket{\theta_1,\varphi_1}{\theta_2,\varphi_2}=\left(\frac{1+\vec{\sigma}_1\cdot \vec{\sigma}_2}{2}\right)^{N},
\end{align}
where $\vec{\sigma}=(\sigma^x,\sigma^y,\sigma^z)$.  From Eq.\eref{eq:overlap} we can easily calculate the probabilities to be in the symmetry-broken ground states $P_{\pm}=\exp(-N f_{\pm})=\left(\frac{1\pm \sx}{2}\right)^{N}.$ Cusps in the return probability $P=P_-+P_+$ occur whenever 
\begin{align}
\label{eq:screturn}
\exp(-f_+)-\exp( -f_-) = \sx=0
\end{align}
Therefore while the presence of cusps in the return probability is a signature of symmetric trajectories, the absence of cusps is a signature of symmetry-broken trajectories, with respect to the transformation $\sigma_x\rightarrow-\sigma_x$. 

The physical meaning of the dynamical phase transition signalled either by the cusps in the return probability or the time-averaged order parameter $\overline{\sx}$ can be understood from \fref{fig3}, where we show different types of trajectories and the energy function in the semiclassical phase-space. If the transverse-field is larger than the zero temperature critical field ($h_f>2$),  we are in the paramagnetic phase where the ground state, the energy function and all trajectories are symmetric. Conversely, if $h_i,h_f<2$ the phase space splits into a symmetry-broken part at low energies and a symmetric part at high energies. The time-averaged order parameter and the cusps in the return probability, indicate a bifurcation of the trajectories as we change the initial or the final Hamiltonian. In contrast to quenches from the ferromagnetic side, the quenches starting in the paramagnetic region do not display signatures of the bifurcation of the phase space. The reason is that the paramagetic ground state is an unstable fixed point of the semiclassical evolution.
\begin{figure}[h!!]
\vskip 0.5truecm
\includegraphics[width=0.9\textwidth]{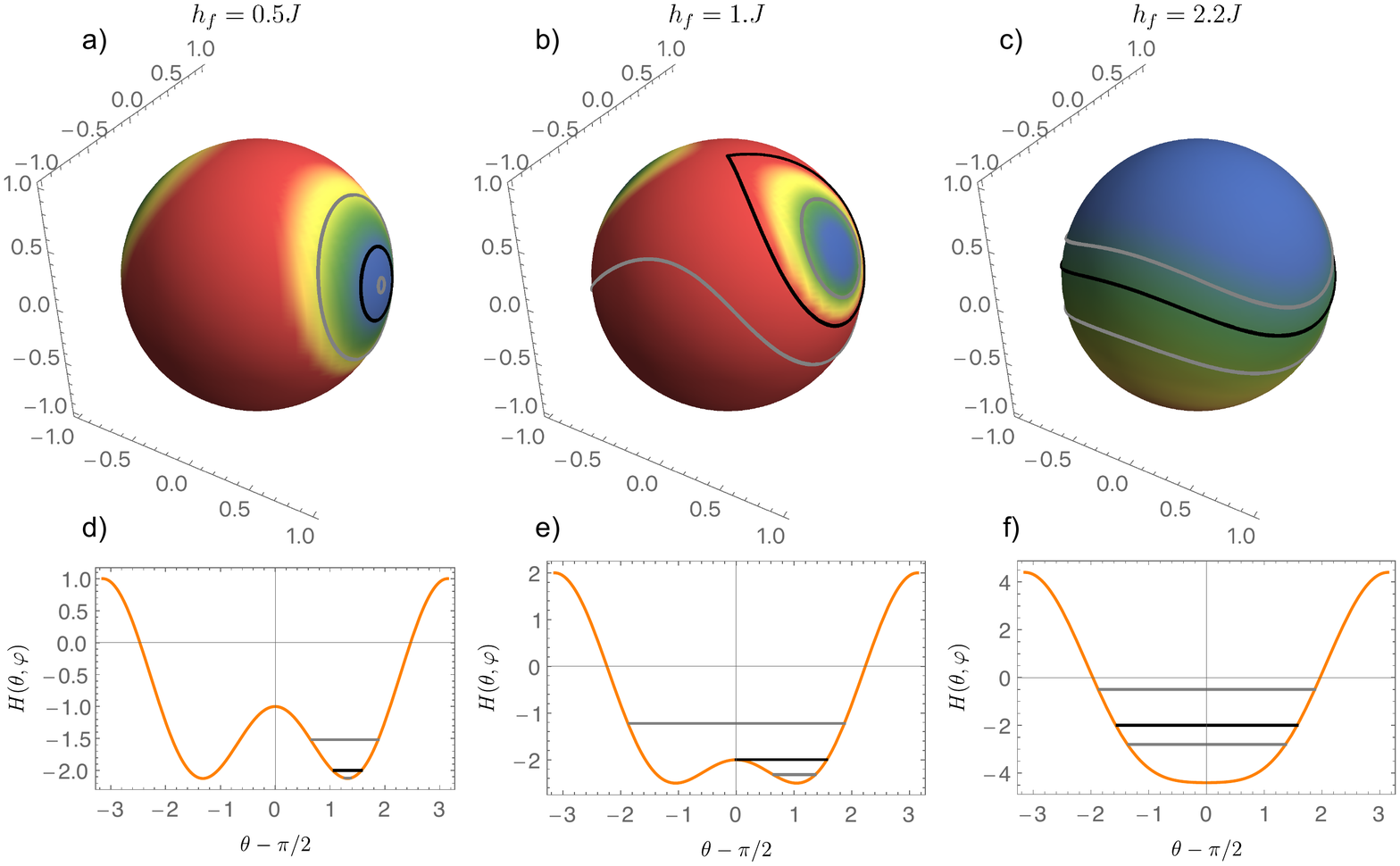}
\caption{Panels (a)-(c) show phase space portraits of the energy function $H(\theta,\varphi)$, where the polar angle (azimuth) $\theta$ is measured from the $xy$ plane and the angular angle $\varphi$ from the $x$ axis. The energy is displayed by color -- we use warm colors for high energy and cold colors for low energy. Panels (d)-(f) show a slice at $\varphi=0$ of the corresponding phase space portrait. The black line corresponds to the quench from $h_i=0$, the gray lines correspond to initial states with a bit higher and lower initial energy with respect to the final Hamiltonian.}
\label{fig3}
\end{figure} 
\newpage

\subsubsection{Finite size scaling}
\begin{figure}[h!!]
\vskip 0.5truecm
\includegraphics[width=0.3\textwidth]{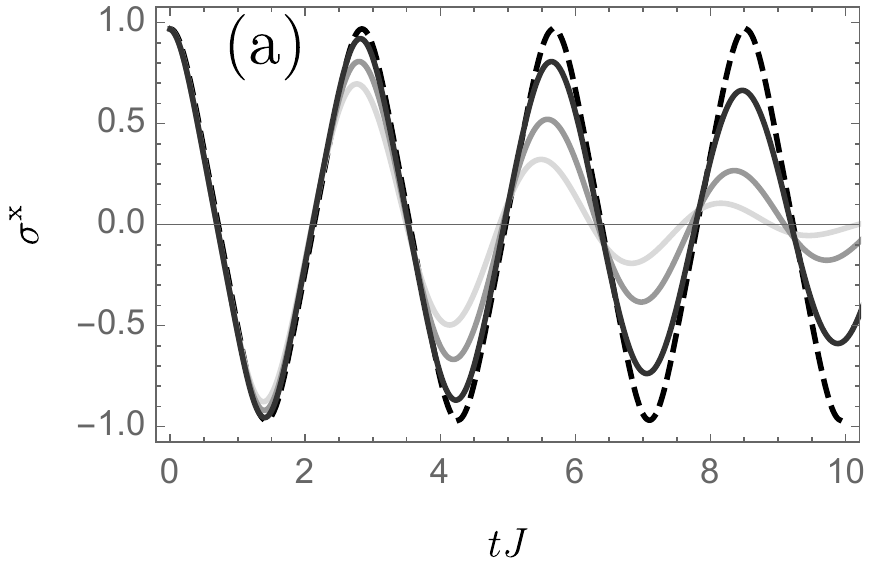}~~\includegraphics[width=0.3\textwidth]{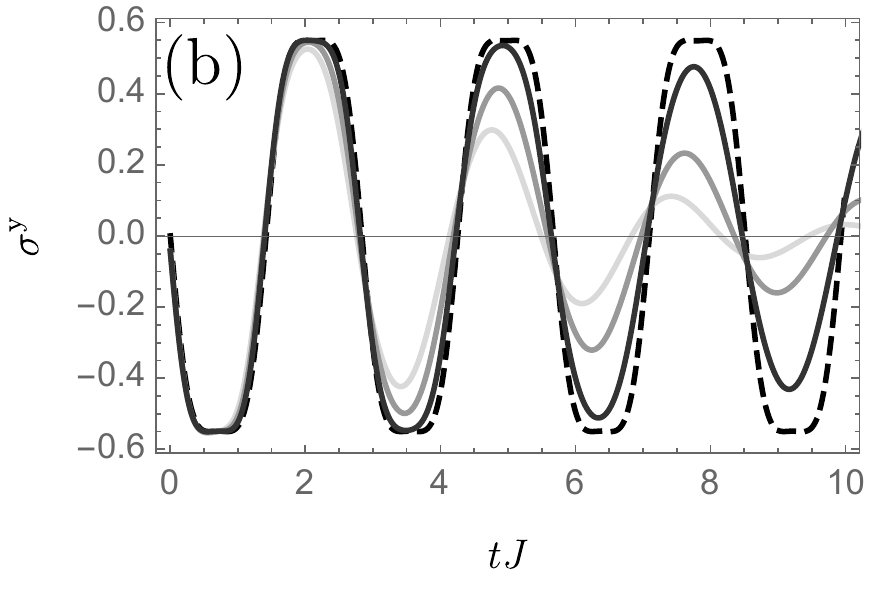}~~\includegraphics[width=0.3\textwidth]{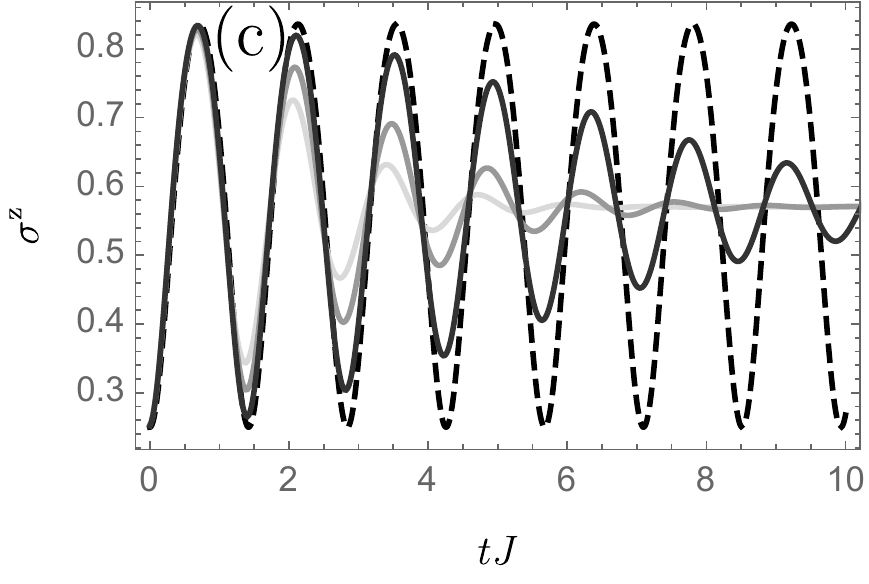}
\includegraphics[width=0.3\textwidth]{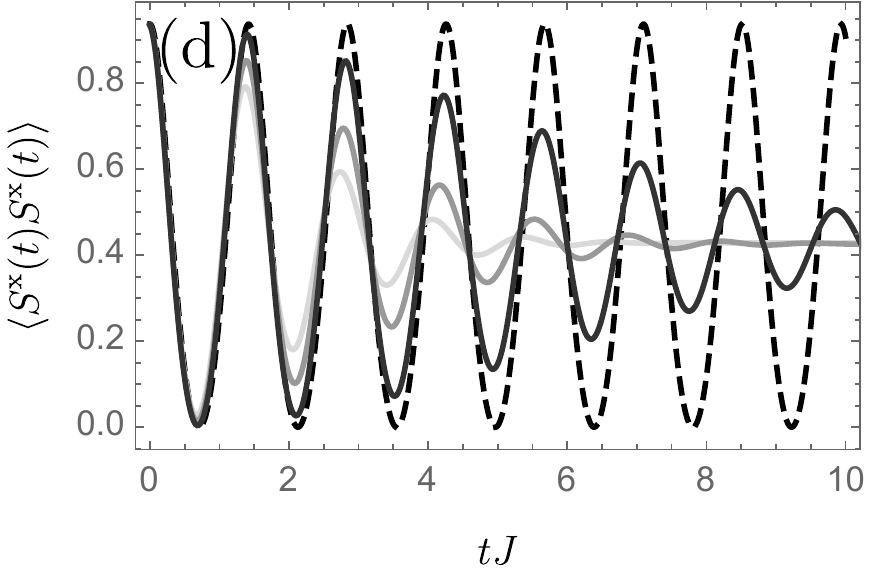}~~\includegraphics[width=0.3\textwidth]{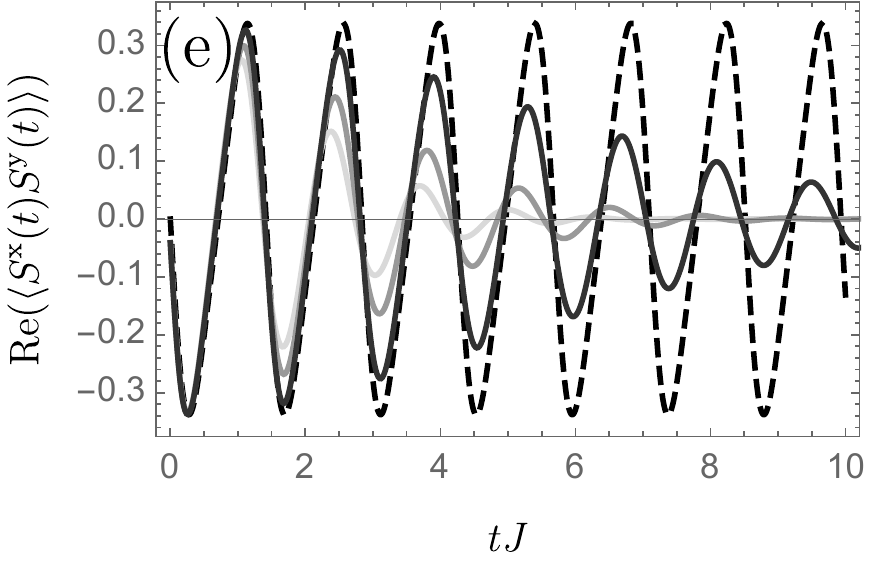}~~\includegraphics[width=0.3\textwidth]{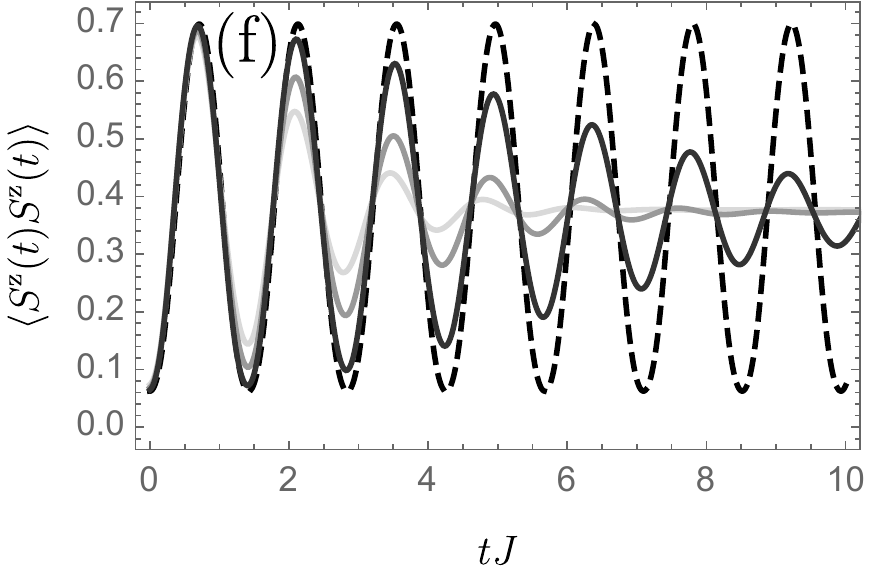}
\caption{Convergence of local observables ( (a)-(c) on-site expectations, (d)-(f) two point correlations ) to the semiclassical values with the increasing system size $N=$100, 200, 800 (from bright to dark). The dashed line shows the semiclassical result. The quench from the feromagnetic phase ($h_i=0.5$) to the paramagnetic phase ($h_f=1.6$) is presented.}
\label{fig8}
\end{figure}

\begin{figure}[h!!]
\vskip 0.5truecm
\includegraphics[width=0.3\textwidth]{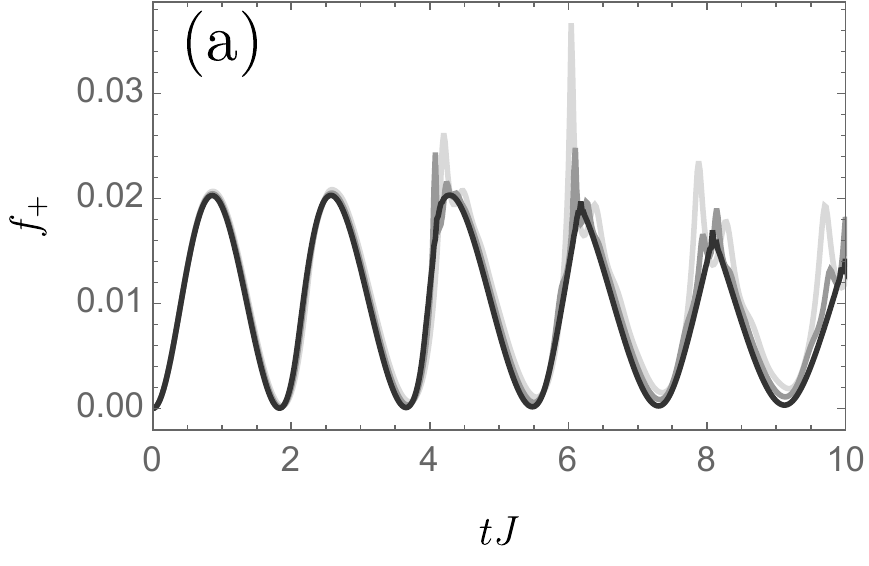}~~\includegraphics[width=0.3\textwidth]{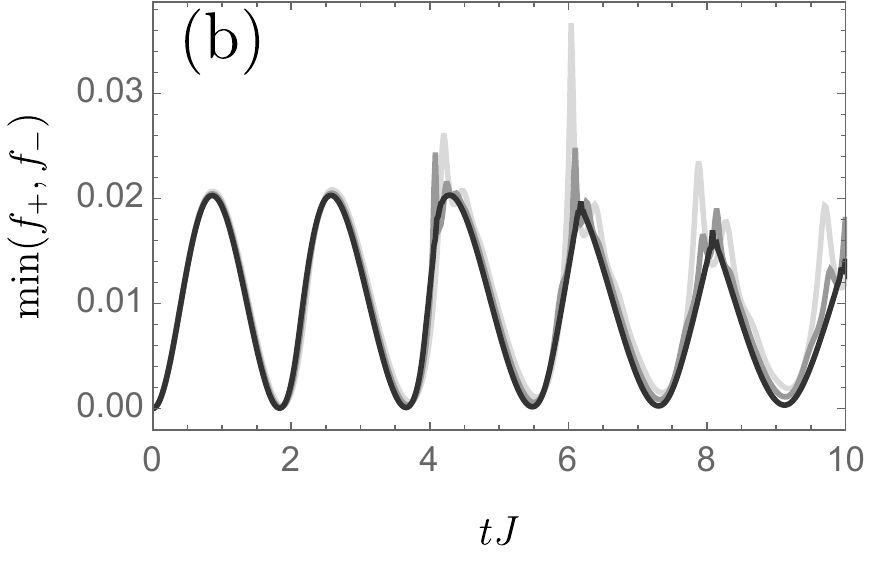}~~\includegraphics[width=0.3\textwidth]{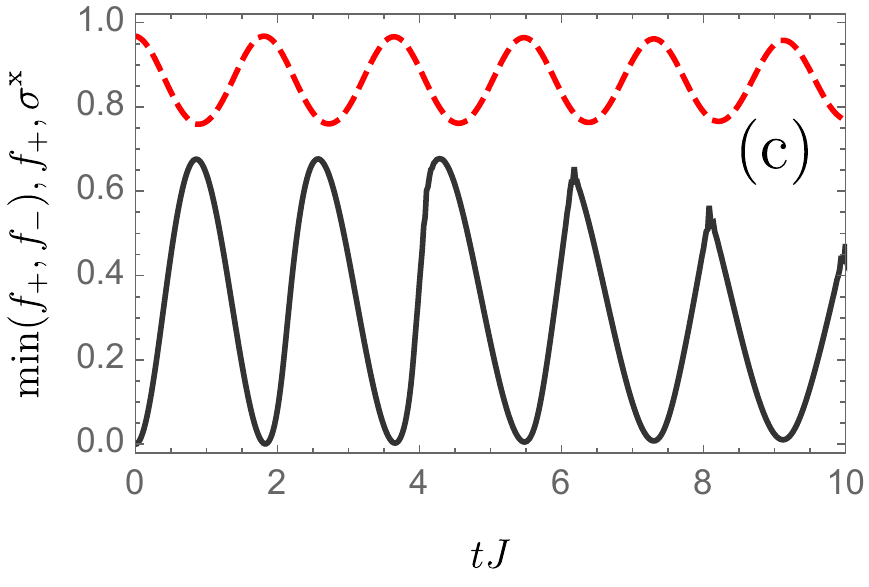}
\includegraphics[width=0.3\textwidth]{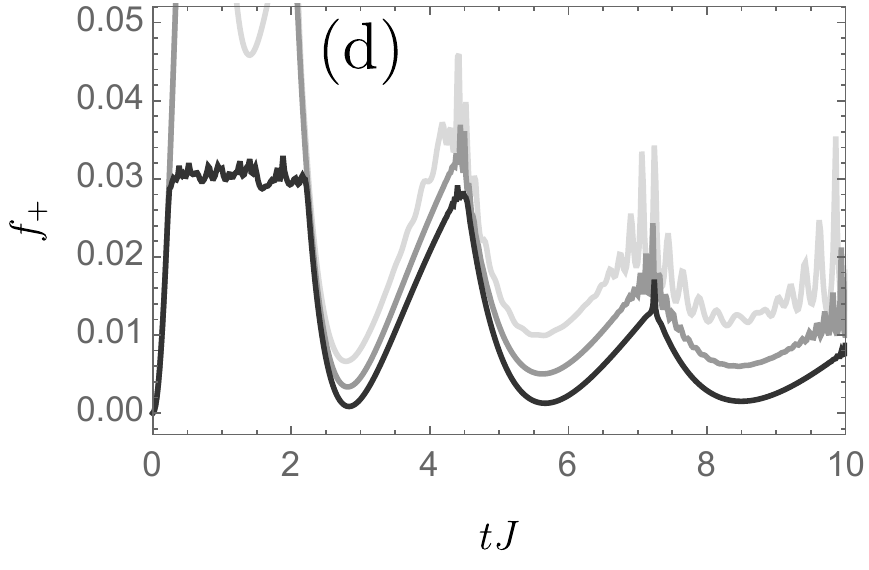}~~\includegraphics[width=0.3\textwidth]{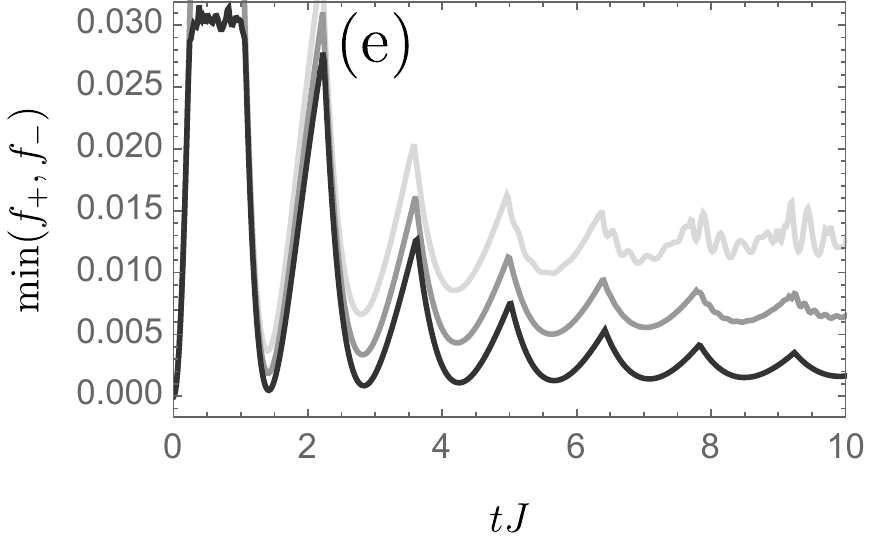}~~\includegraphics[width=0.3\textwidth]{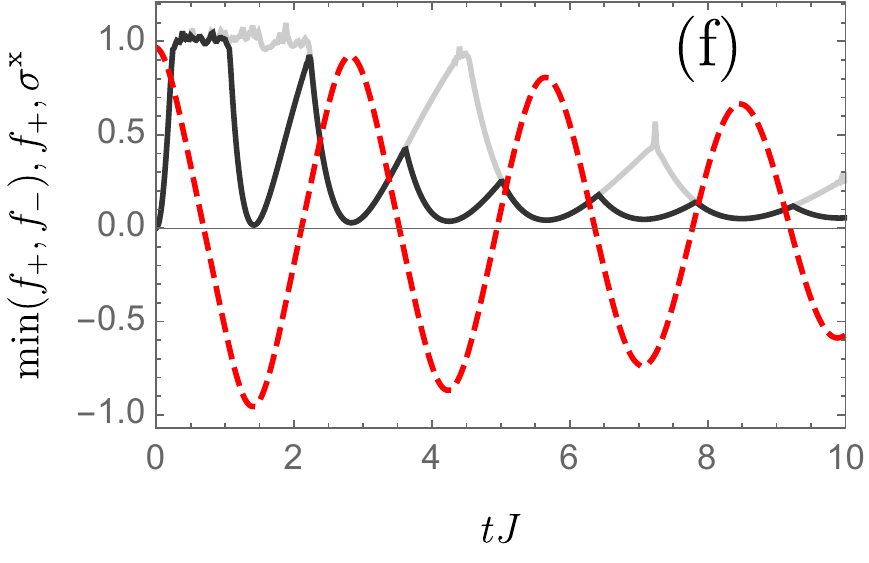}
\caption{Finite size scaling of the return probability (b,e) and the Loschmidt echo  (a,d) for a quenches from $h_i=0$ to $h_f=0.9$ (a,b,c) and $h_f=1.6$ (d,e,f). If we stay on the ferromagnetic side the singularities that may appear for finite system sizes are smoothened. Conversely, for the quenches to the paramagnetic side ($h=1.6$) the singularities become more pronounced as the system size is increased. Similarly for the return probability we observe singularities only for quenches across the dynamical critical point. In panels (c,f) we compare the rescaled Loashmidt echo (gray line) and the return probability (black line) with the time evolutions of the longitudinal magnetisation $\sx(t)$ (red dashed line). We observe that as in the semiclassical case the singularities in the Loschmidt echo correspond to the minima of $\sx(t)$, whereas the singularities in the return probability correspond zeros of $\sx(t)$. Accordingly, the period of cusps in the Loschmidt echo is twice the period of the cusps in the return probability. On the first two plots we show data system sizes $N=100, 200, 800$ from bright to dark. The right plots show rescaled data for $N=800$.}
\label{fig3new}
\end{figure}

While the semiclassical formula Eq.(\ref{eq:overlap}) predicts correctly the cusps emerging in $P(t)$ when $P_+=P_-$, it should be stressed that also the individual contributions $P_+,P_-$ may have singularities in time. These are correctly predicted by Eq.(\ref{eq:overlap}) only for $h_i=0$ while for $h_i\neq 0$ semiclassics smoothens artificially such singularities. However, it is important to stress that, as explained below, these additional singularities are not observed in $P(t)$ which is therefore correctly described by the semiclassical formula.  Let us discuss this point more in depth by performing a careful finite size scaling of the model.  To obtain exact results for finite size systems we diagonalise the Hamiltonian Eq.\eref{eq:hbigspin} in the spin $N/2$ subspace with the two degenerate ground states (with exponentially close energy as $N$ increases). We first show that by increasing system size the trajectory traced out by the magnetisation converges to the semiclassical value (which is also true for higher moments; see \fref{fig8}).
In contrast to the local observables, the finite size scaling of the return probabilities and the Loschmidt echo approaches the semiclassical result only in part as the system size is increased, since the latter is not able to resolve the time singularities present in the exact solution: The Loschmidt echo computed exactly displays cusps at the minima of $\sx(t)$ (for all system sizes), which are in turn absent in the semiclassical description for quenches from $h_i>0$. This is essentially due to the fact that the Loschmidt echo {\it is not} a local observable.  In turn singularities in $P(t)$ are correctly predicted by the semiclassical formula: this is  shown in \fref{fig3new} where the singularities in the Loschmidt echo align with the minima of the order parameter $\sx(t)$, whereas the singularities of the return probability align with the zeros of the order parameter $\sx(t)$. Since, when performing the finite size scaling, the trajectory traced out by the total magnetisation approaches the semiclassical description (see \fref{fig8}) the interpretation of the Loschmidt echo and the return probability as signatures of bifurcation of the trajectories robust.

Let us now briefly discuss quenches from the paramagnetic side. In this case, the time-averaged order parameter remains zero at all times while fluctuations are non-vanishing (if the system size is finite). In \fref{fig4} we show the time average of the fluctuations as a function of the post-quench magnetic field for different system sizes. We observe a change of behaviour at the quantum critical point $h_c=2$. As we approach the thermodynamic limit we see that the signature of the quantum phase transition vanishes algebraically with the system size (see \fref{fig4} right). We observe similar behaviour of the return probability. Since the fully polarised state in the transverse direction becomes an unstable fixed point of the semiclassical dynamics the ground state probability remains constant. The finite size scaling reveals an exponential decay of the Loschmidt amplitude with time for quenches to the ferromagnetic phase shown in \fref{fig4a} for a quench from a fully polarised state ($h_i\rightarrow\infty$) to $h_f=1/2$ for different system sizes.
\begin{figure}[h!!]
\vskip 0.5truecm
\includegraphics[width=0.4\textwidth]{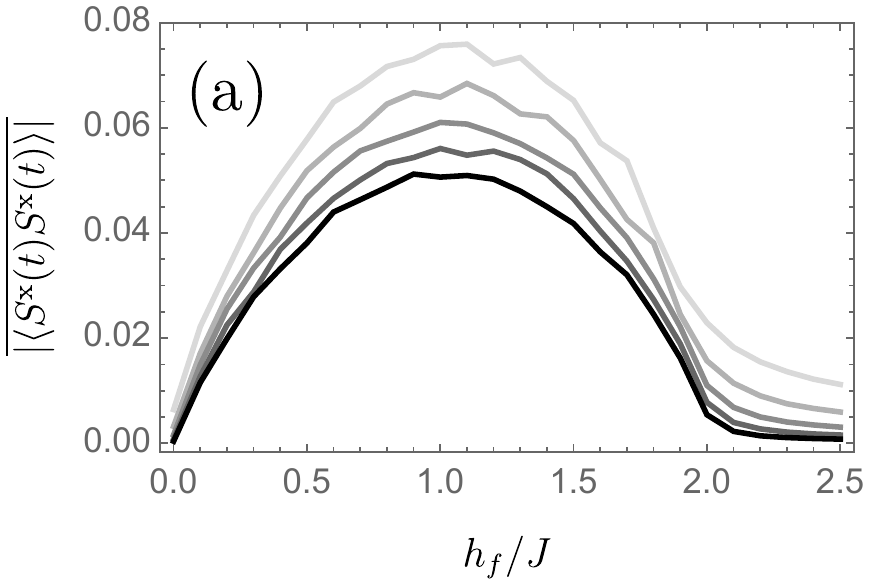}~~\includegraphics[width=0.43\textwidth]{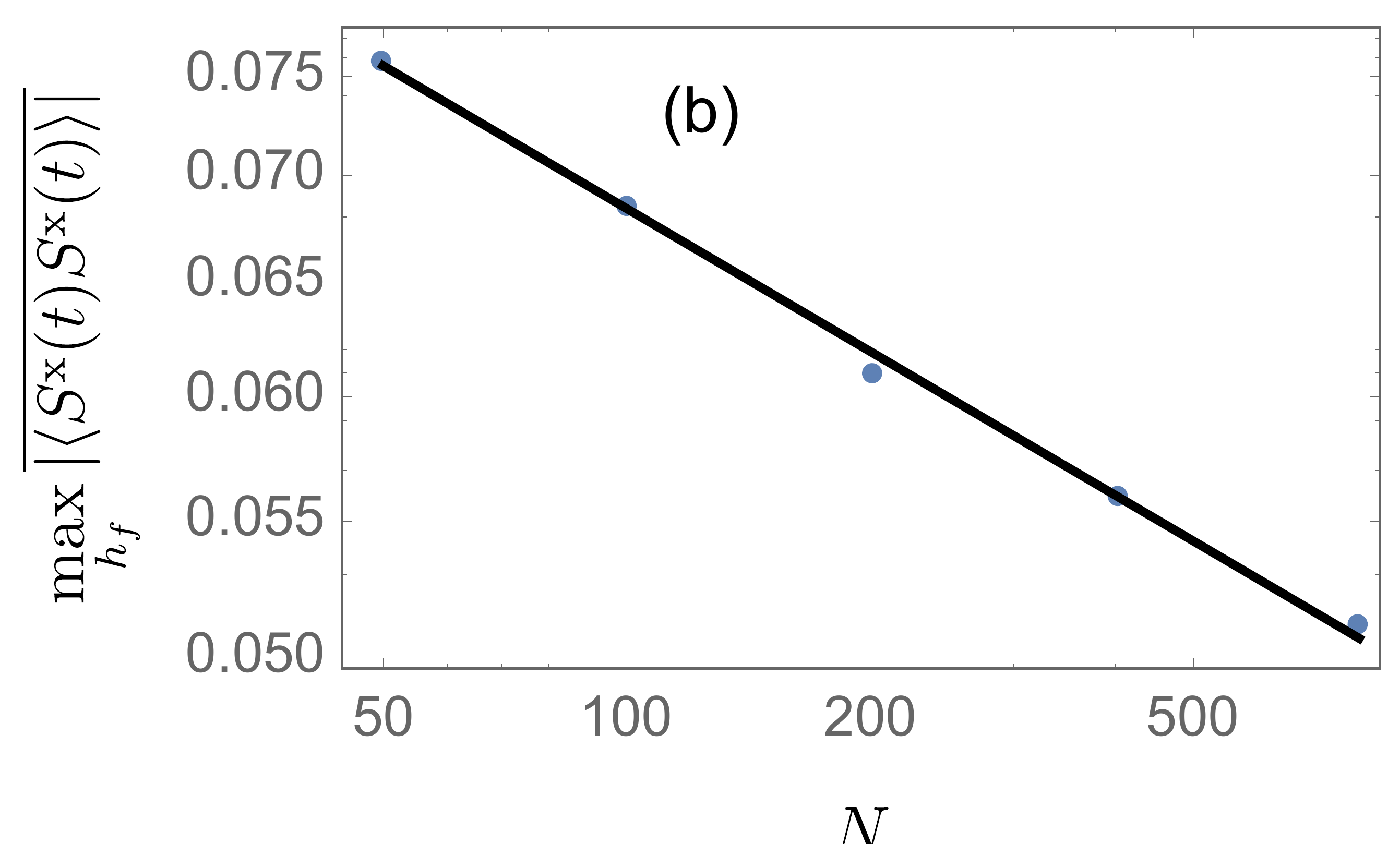}
\caption{(a) Fluctuations as a function of the postquench magnetic field for different system sizes. (b) Dependence of the maximum of  $\overline{\ave{S_X^2}}$ on the system size $N=50, 100, 200, 400, 800$ (from bright to dark).}
\label{fig4}
\end{figure}
\begin{figure}[h!!]
\vskip 0.5truecm
\includegraphics[width=0.4\textwidth]{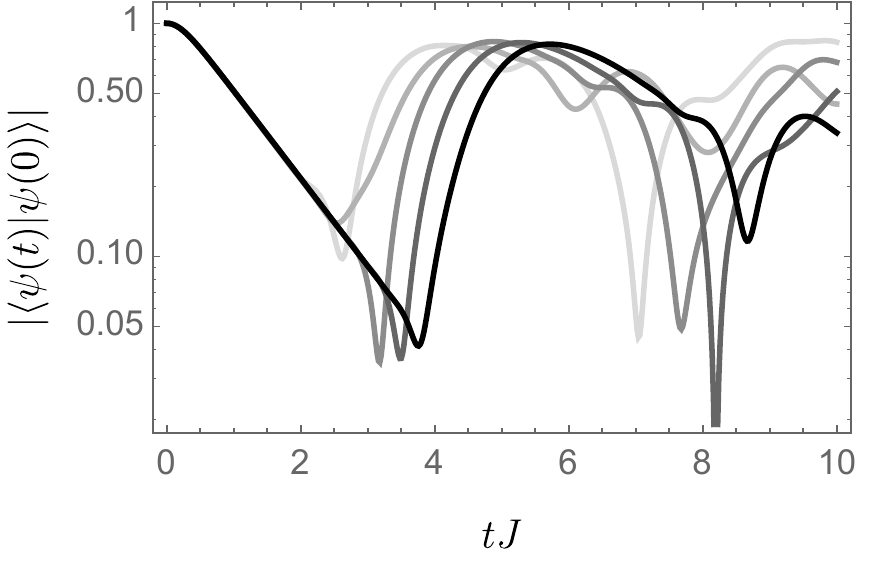}
\caption{Finite size scaling of the return probability (log-linear scale). We start with a fully polarised state and evolve with the final Hamiltonian in the ferromagnetic phase ($h_f=1/2$).  The colors denote the system size  $N=50, 100, 200, 400, 800$ (from bright to dark).}
\label{fig4a}
\end{figure} 

\subsubsection{Difference between dynamical and thermal phase transitions}
\begin{figure}[h!!]
\vskip 0.5truecm
\includegraphics[width=0.305\textwidth]{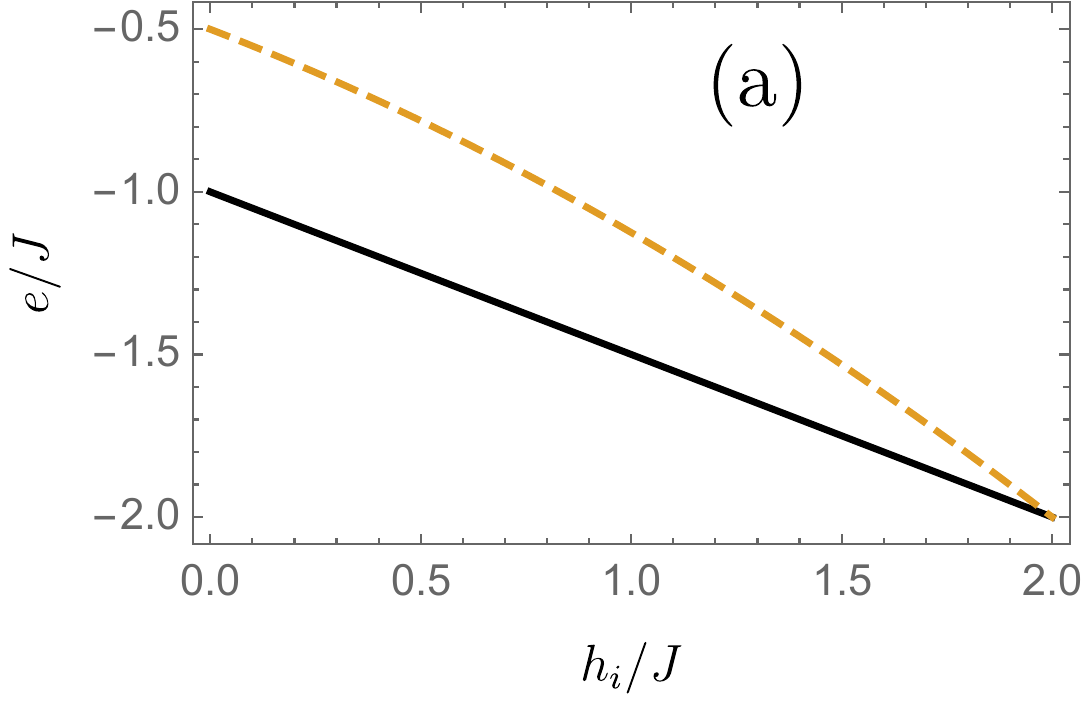}~~\includegraphics[width=0.3\textwidth]{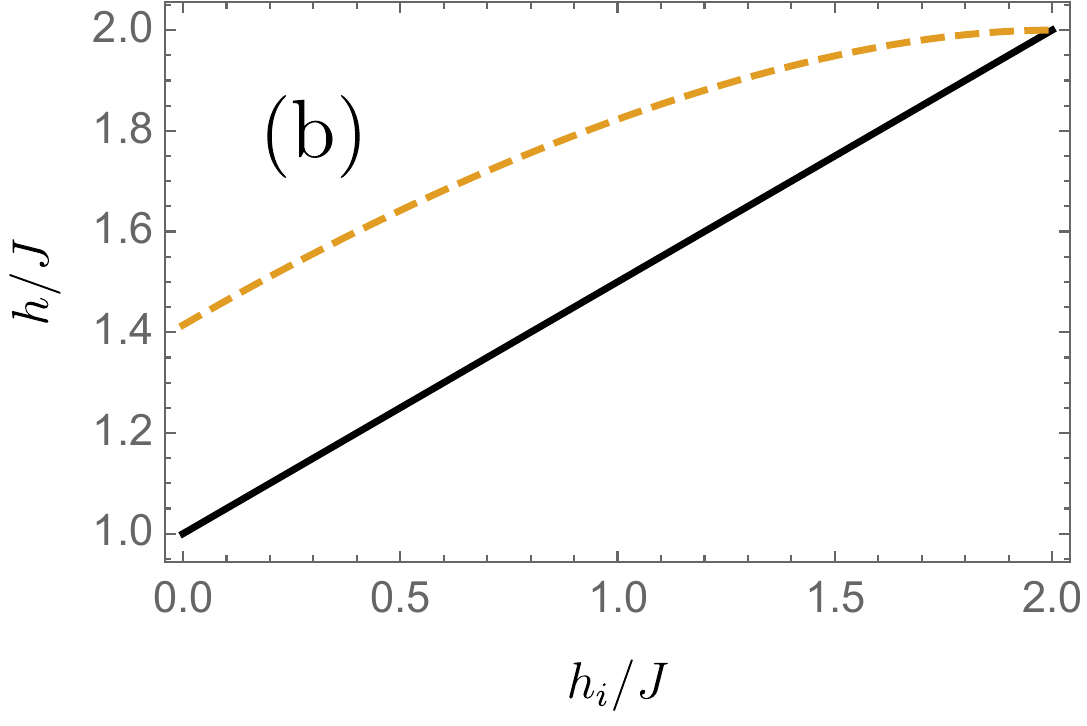}
\caption{(a) Energy densities $e_1$ (black line) and $e_2$ (orange dashed) as a function of the initial magnetic field. (b) The dynamical critical field (black line) and the equation \eref{eq:hfe1e2} (orange dashed) as a function of the initial magnetic field. In both cases we see significant difference for small initial magnetic fields.}
\label{fig5}
\end{figure}

In order to asses if there is any relation between the dynamical and the thermal phase transition in the infinite-range Ising model we compare the energy density of the critical trajectory governing the phase transition at some $h_f$ with the energy density associated to the critical temperature at the same $h_f$. 
We start by writing the energy of the initial state (the ground state of the initial Hamiltonian corresponding to $h_i$) with respect to the final Hamiltonian
\begin{align}
\label{eq:ehh}
e(h_i,h_f)=\frac{-2 h_f h_i + h_i^2 - 4 J^2}{4J}.
\end{align}
At the critical field $h_f=h_c=J+h_i/2$ we obtain the energy density of the critical trajectory corresponding to the quench from $h_i$
\begin{align}
e_1=-J-\frac{h_i}{2}.
\end{align}
We compare this energy density with the one corresponding to the critical temperature at $h_f=h_c$.  From the Eq.\eref{eq:Tc} we calculate the critical temperature $T_c(h_c)=h_c/{\rm arctanh}(h_c/2J)$, which gives the one particle mean-field density matrix at the critical point
\begin{align}
\rho_{T_c}=\frac{1}{2}\begin{pmatrix}
\frac{1}{2}+\frac{h_c}{4J}&0\\
0&\frac{1}{2}-\frac{h_c}{4J}
\end{pmatrix}.
\end{align}
From this we get the thermal energy density
\begin{align}
e_2=-\frac{h_c^2}{2J}=-\frac{(h_i+2J)^2}{8J}.
\end{align}
From \fref{fig5} left, where we plot both energy densities, we see that the energy density of the critical trajectory is strictly smaller than the thermal density. 
The final magnetic field at which the energy densities corresponding to the thermal and initial state coincide is
\begin{align}
\label{eq:hfe1e2}
h_f(e_1=e_2)=\frac{1}{2} \left(\sqrt{8 J^2-h_i^2}+h_i\right).
\end{align}
In \fref{fig5} right we compare the dynamical critical field with the equation \eref{eq:hfe1e2} and observe that the field strength, at which the thermal energy density agrees with the energy density of the initial state, is significantly larger (for small $h_i$) as the dynamical critical value.

\newpage
\section{Additional results for finite range interactions $\alpha>0$}
In this section we present additional results for the scaling of the time-averaged magnetisation $\overline{\sx}$, the cusps in the Loschmidt echo and the return probabilities in the case of finite interaction ranges.
\subsection{Time-averaged order parameter}
\begin{figure}[h!!]
\vskip 0.5truecm
\includegraphics[width=0.4\textwidth]{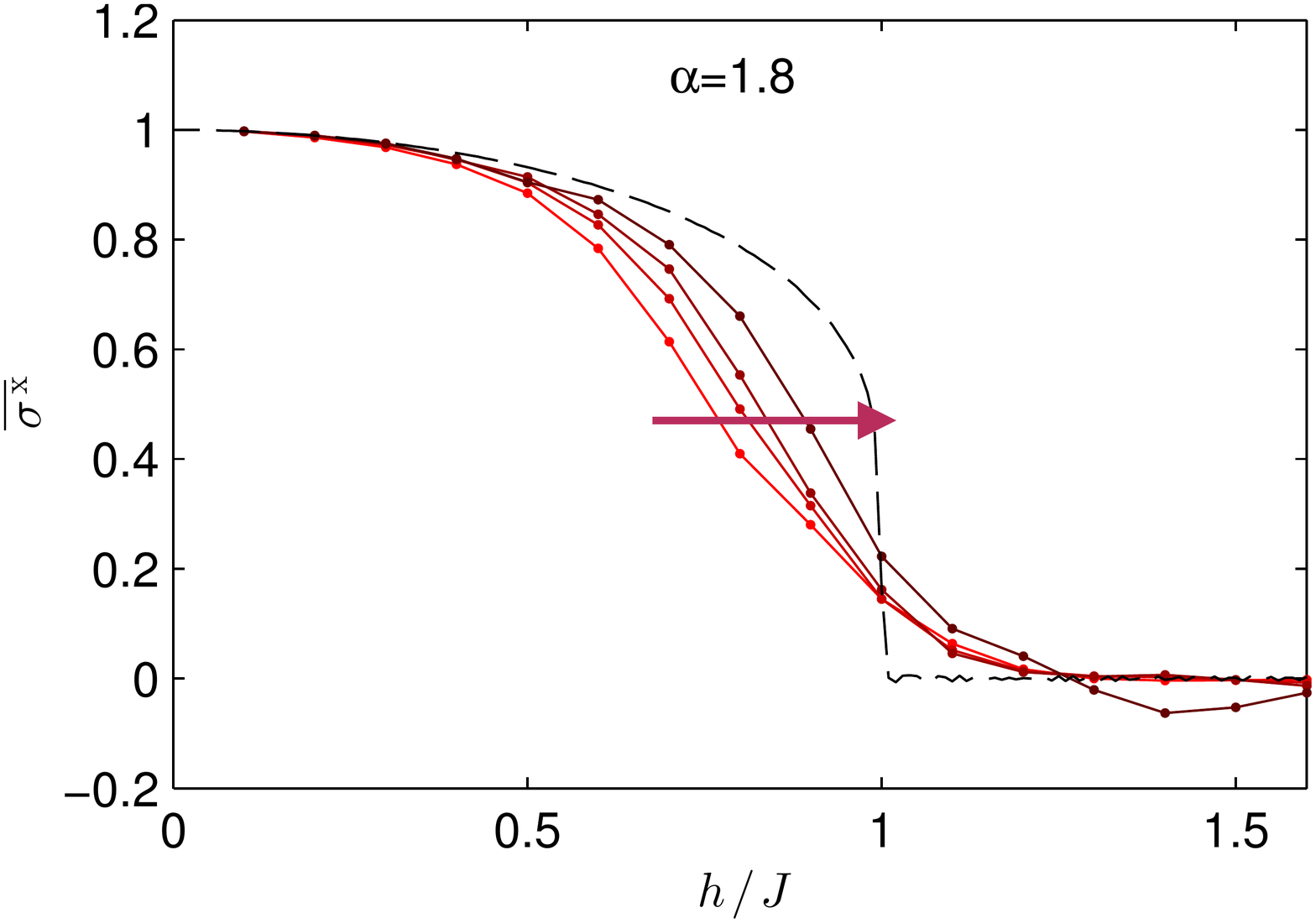}\includegraphics[width=0.4\textwidth]{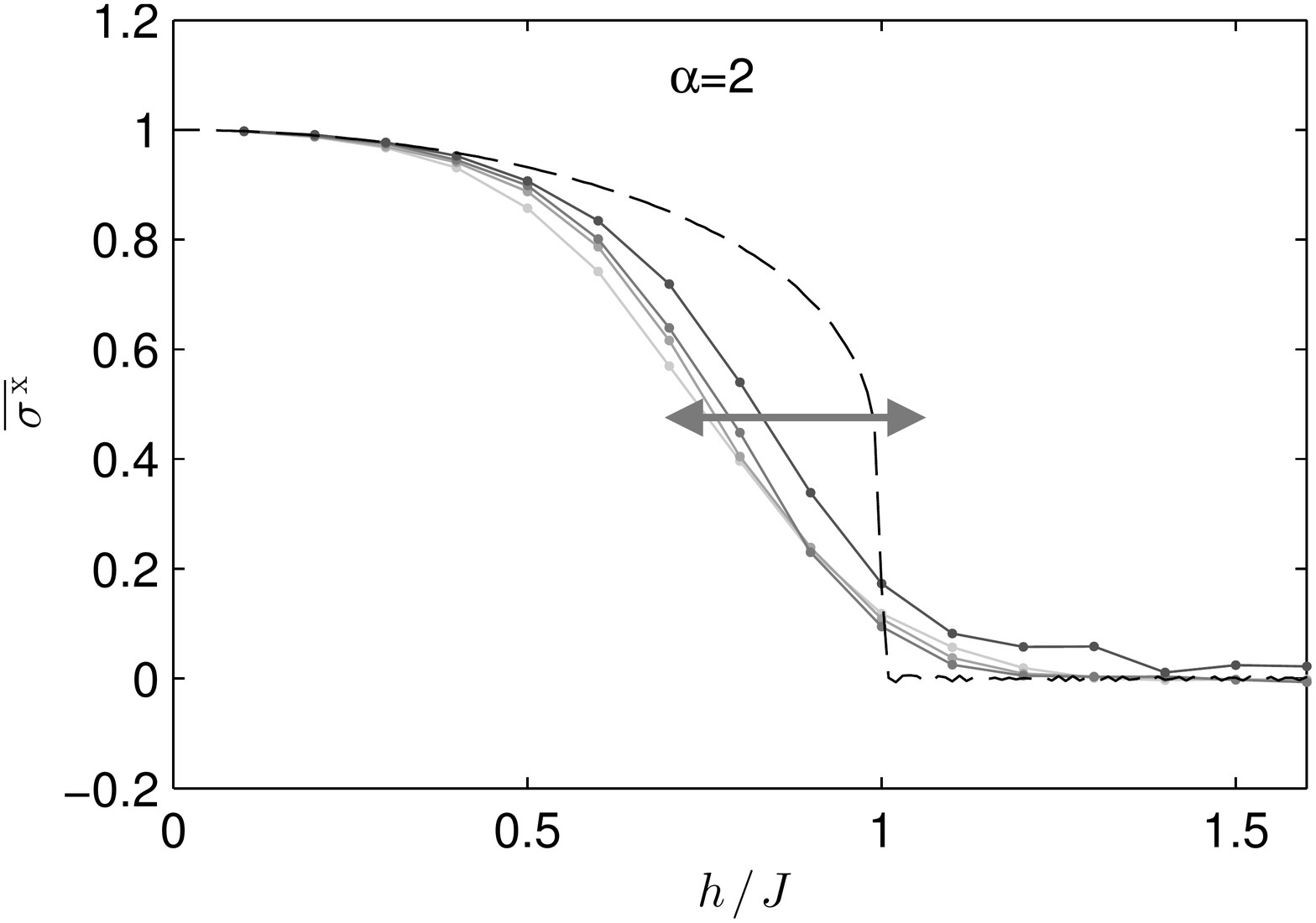}
\includegraphics[width=0.4\textwidth]{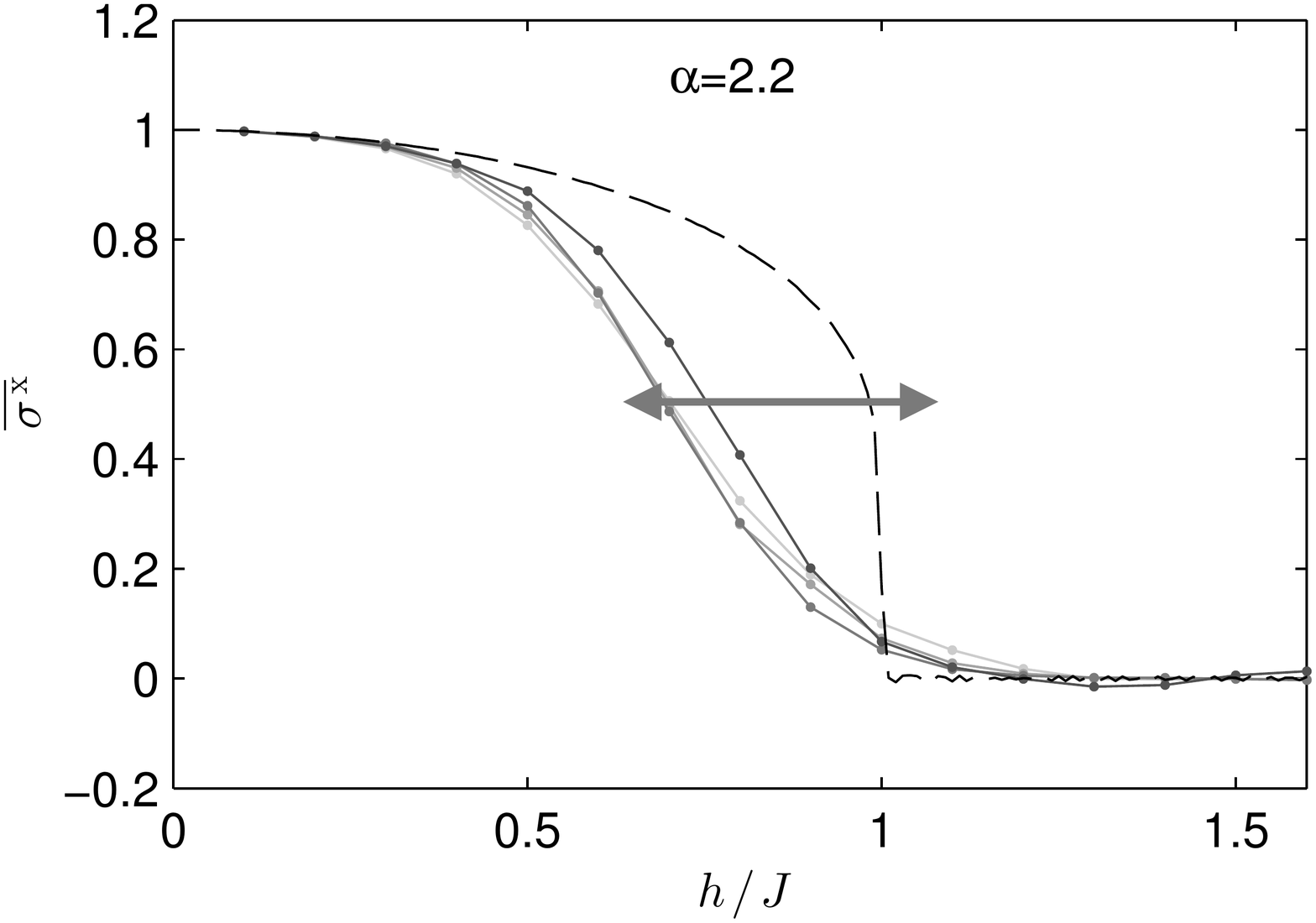}\includegraphics[width=0.4\textwidth]{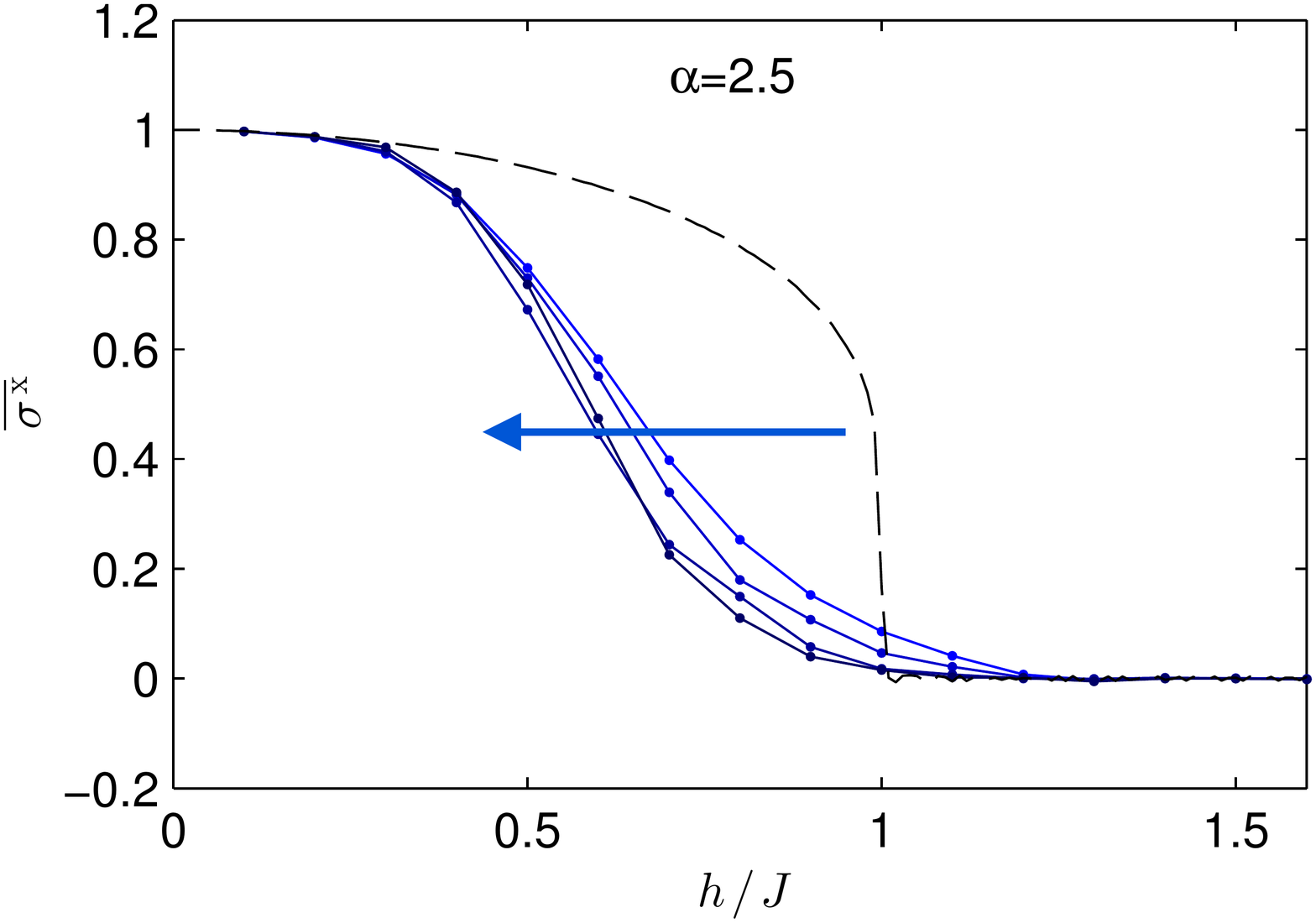}
\caption{Scaling of the time-averaged order parameter $\overline{\sx}$ with the system size. From bright to dark the colours correspond to increasing system sizes $N= 25, 50, 100, 200$.  The plot calculated for $\alpha=1.8$ shows convergence of $\overline{\sx}$ to non-vanishing values for $h<1$, while the plot calculated for $\alpha=2.5$ indicates decrease of $\overline{\sx}$ with the system size for all $h$. The data for $\alpha=2.0$ and $\alpha=2.2$ is inconclusive. The reasons are probably to short simulation times. While for $N=20, 50, 100$ the order parameter is converged, for the largest system size $N=200$ the simulation times were too short in for some data-points.}
\label{fig7a}
\end{figure}

Let us first focus again on the scaling of the time-averaged order parameter with the system size. In \fref{fig7a} we show the scaling of $\overline{\sx}$ with the system size and for $\alpha=1.8$, 2, 2.2 and  $2.5$. For the first, $\alpha=1.8$, we see consistent increase and for the last, $\alpha=2.5$, decrease of $\overline{\sx}$ with the system size. This indicates that in the thermodynamic limit the order parameter stays finite up to some critical field in the former case whereas it vanishes for all values of the field in the latter case. Although, the simulation results for $\alpha=2$ and $\alpha=2.2$ do not provide a clear understanding of the  the large $n$ behaviour of the time-averaged order parameter $\overline{\sx}$, we can confine the transition between vanishing and finite magnetisation to the region $2\lesssim\alpha\lesssim2.4$.

\subsection{Loschmidt echo and return probability}
\begin{figure}[h!!]
\vskip 0.5truecm
\includegraphics[width=0.4\textwidth]{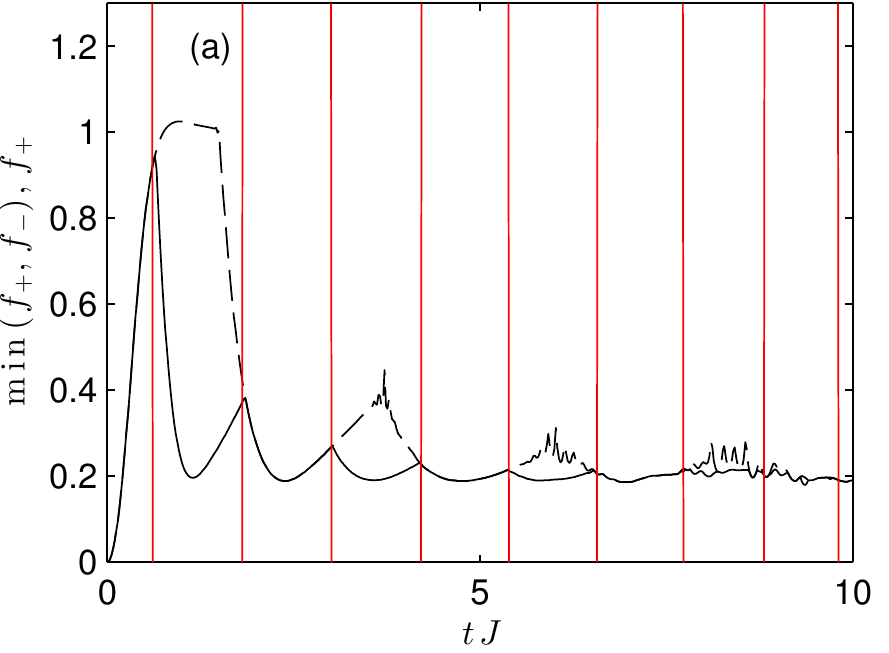}\includegraphics[width=0.4\textwidth]{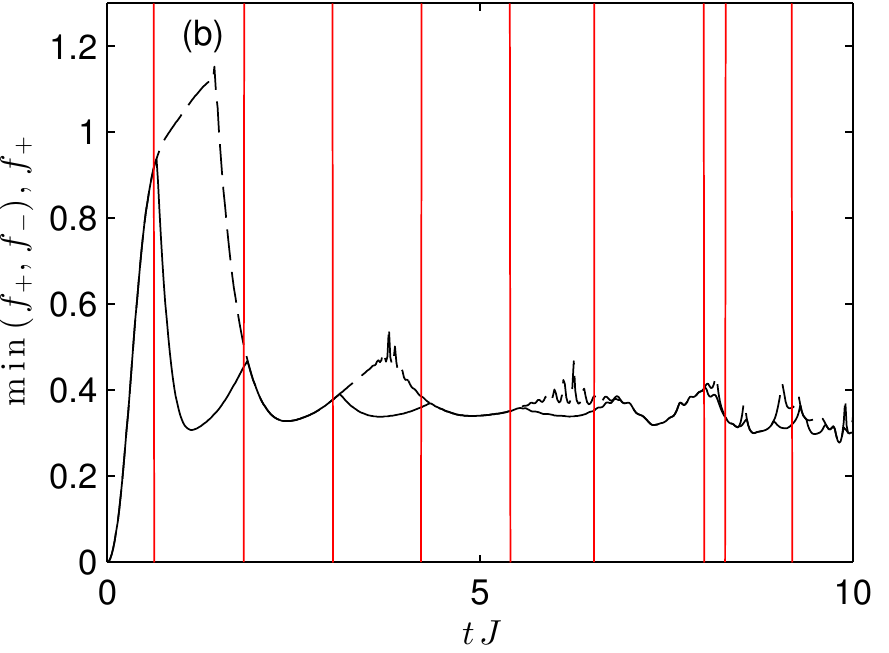}
\caption{Position of the cusp singularities relative of the Loschmidt echo (dashed line) and the return probability (full line) to the zeros of the order parameter $\sx(t)$, indicated by vertical red lines. We show data for $\alpha=1.8$ (a) and $\alpha=2.5$ (b) and observe that the semiclassical picture remains valid also in the case of short range interactions $\alpha>2$, where the time-averaged order parameter vanishes. The system size is 100.}
\label{fig11}
\end{figure}

\begin{figure}[h!!]
\vskip 0.5truecm
\includegraphics[width=0.4\textwidth]{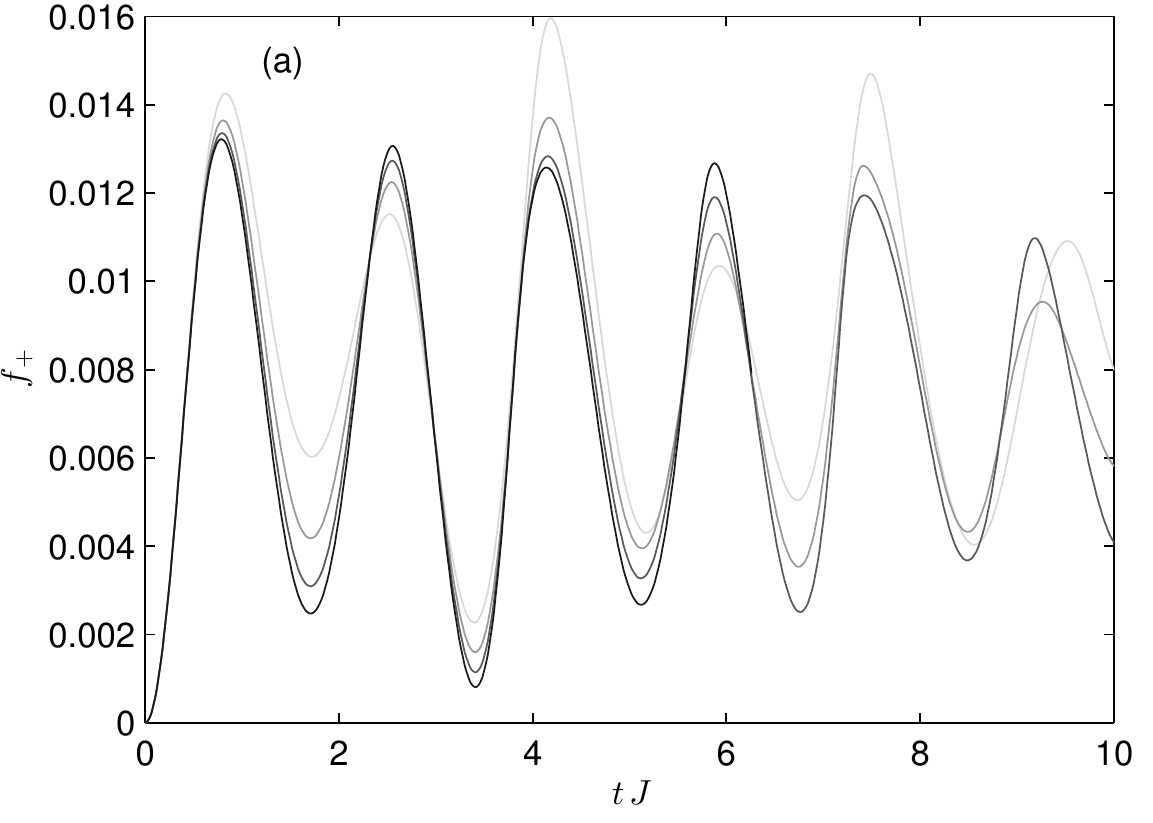}~~\includegraphics[width=0.4\textwidth]{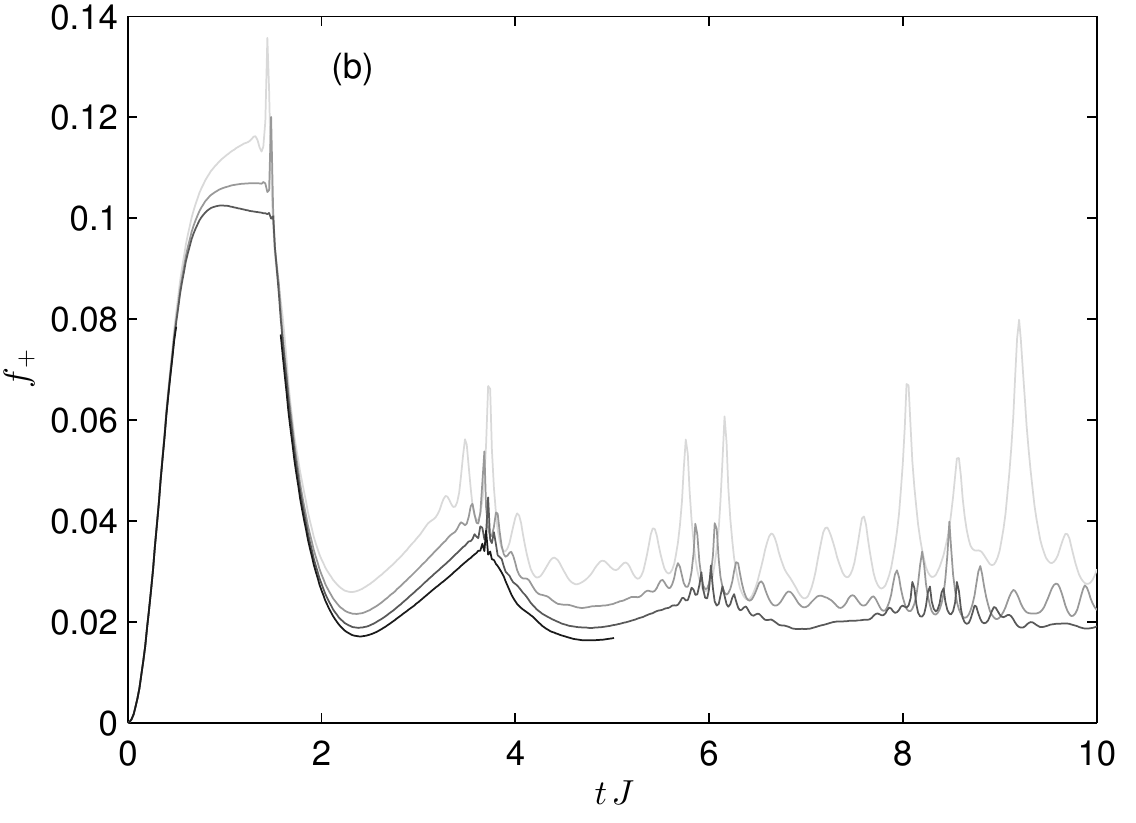}
\includegraphics[width=0.4\textwidth]{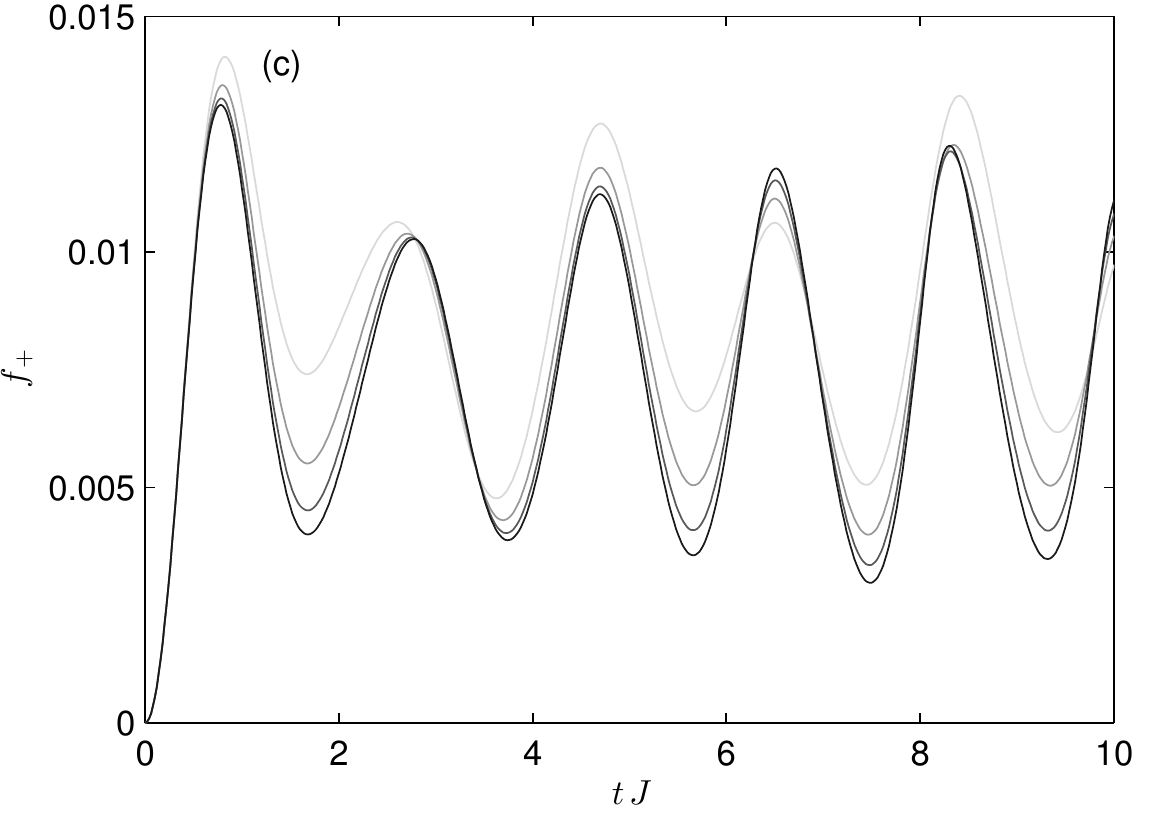}~~\includegraphics[width=0.4\textwidth]{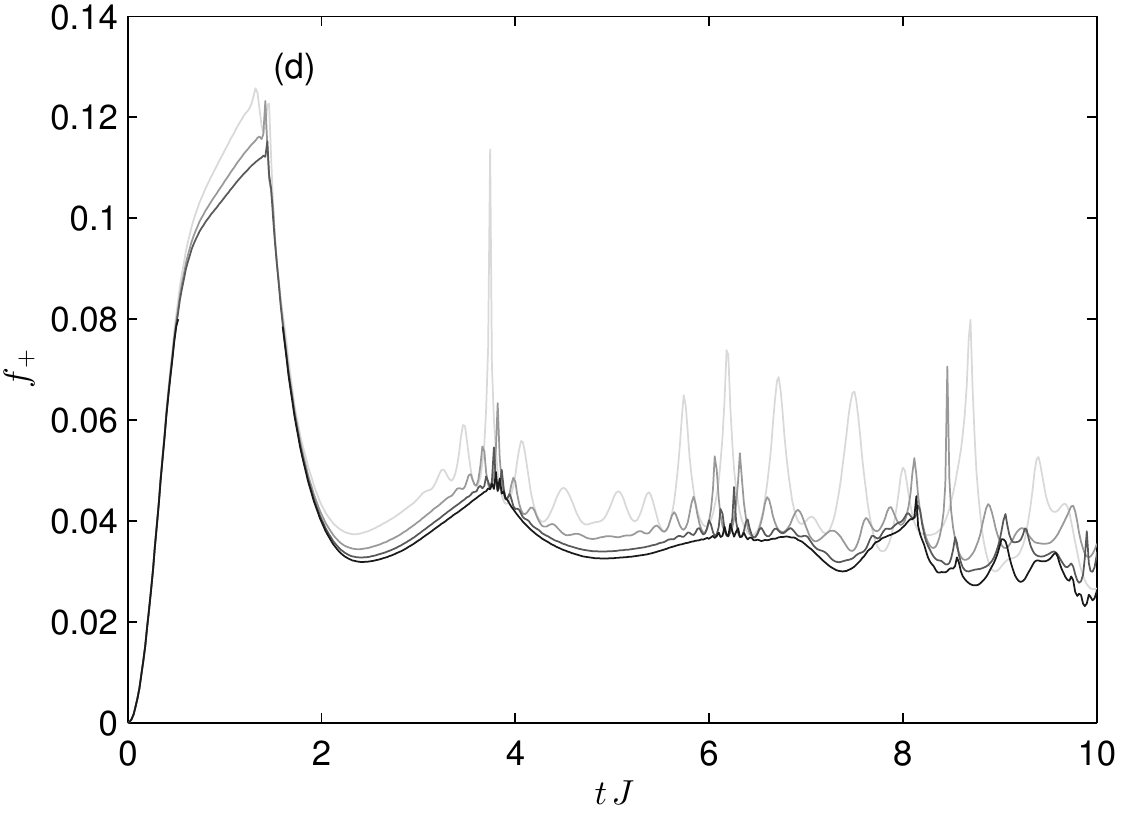}
\caption{Scaling of the Loschmidt echo with the system size for quenches to $h_f=0.5$ (a,c) and $h_f=1.5$ (b,d). The top plots correspond to the interaction range $\alpha=1.8$, where  we still see a non-vanishing time-averaged order parameter. The bottom plots show simulations for $\alpha=2.5,$ where the time-averaged order parameter vanishes, but the cusp singularities  in the Loschmidt echo still persist. The colors correspond to different system sizes; $N=25, 50, 100, 200$ from bright to dark.}
\label{fig9}
\end{figure}

\begin{figure}[h!!]
\vskip 0.5truecm
\includegraphics[width=0.4\textwidth]{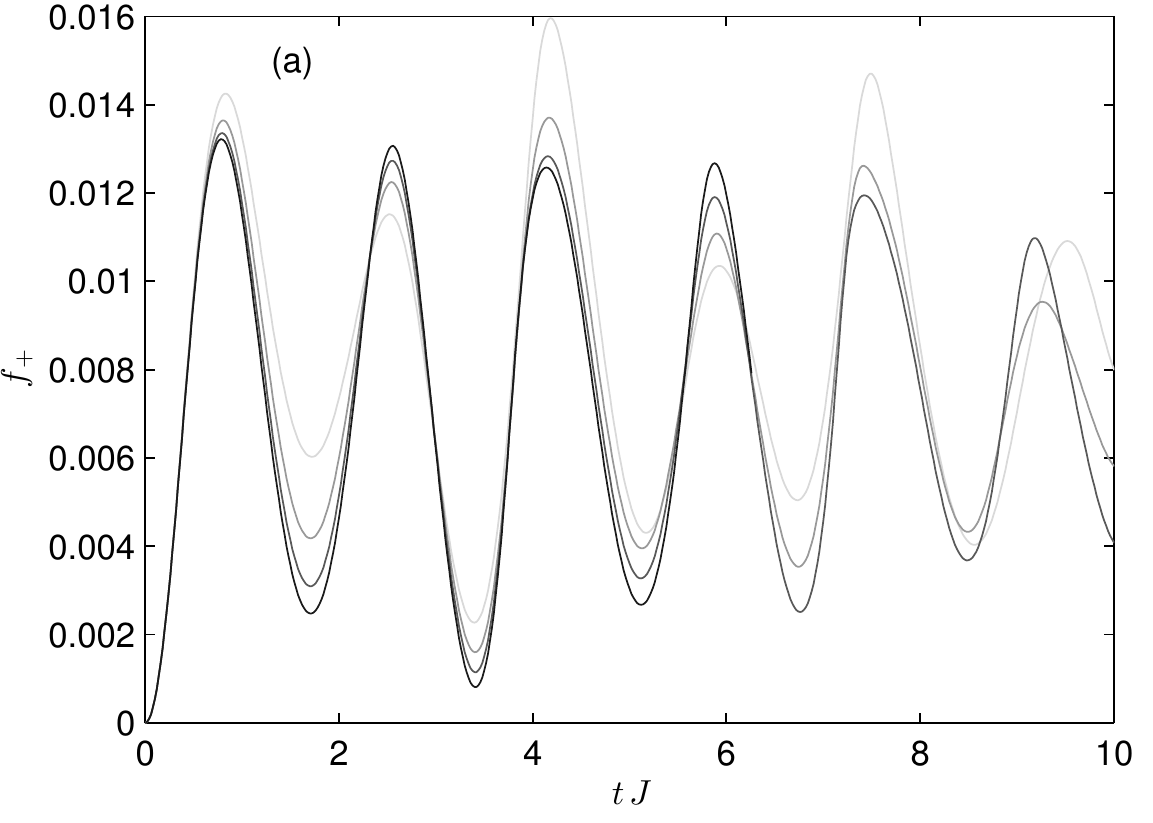}~~\includegraphics[width=0.4\textwidth]{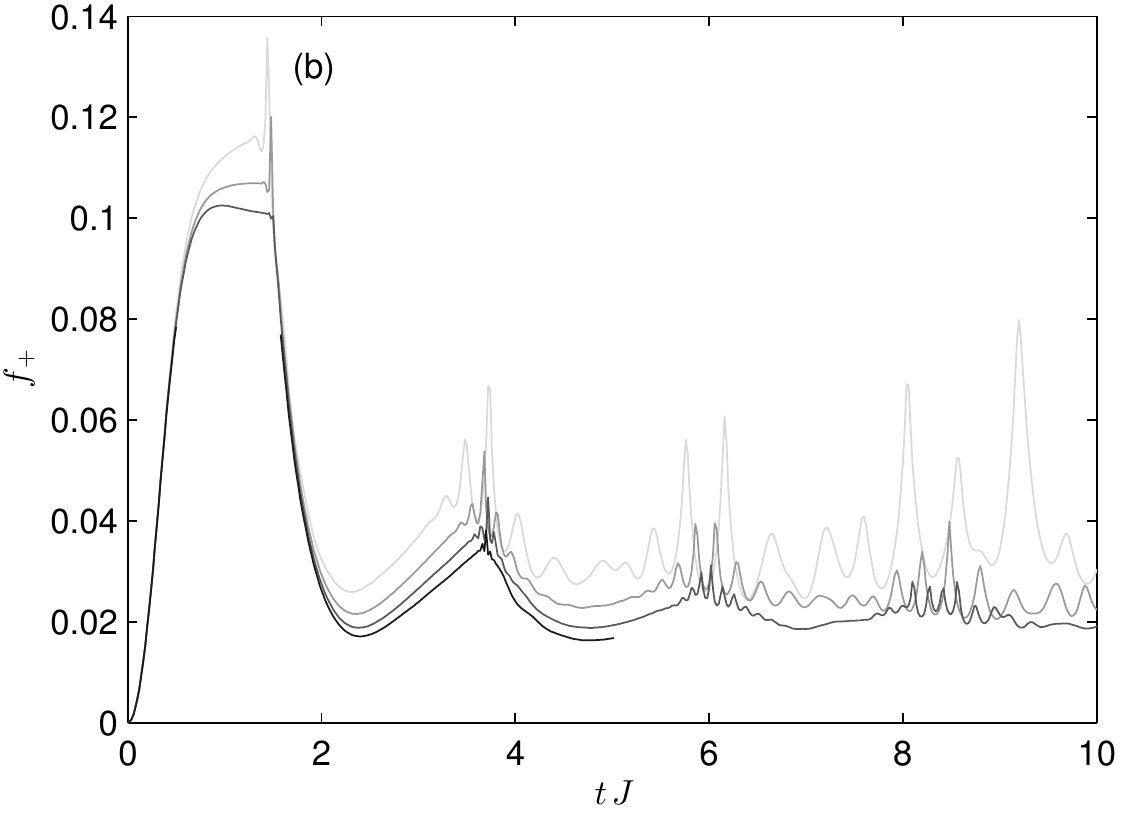}
\includegraphics[width=0.4\textwidth]{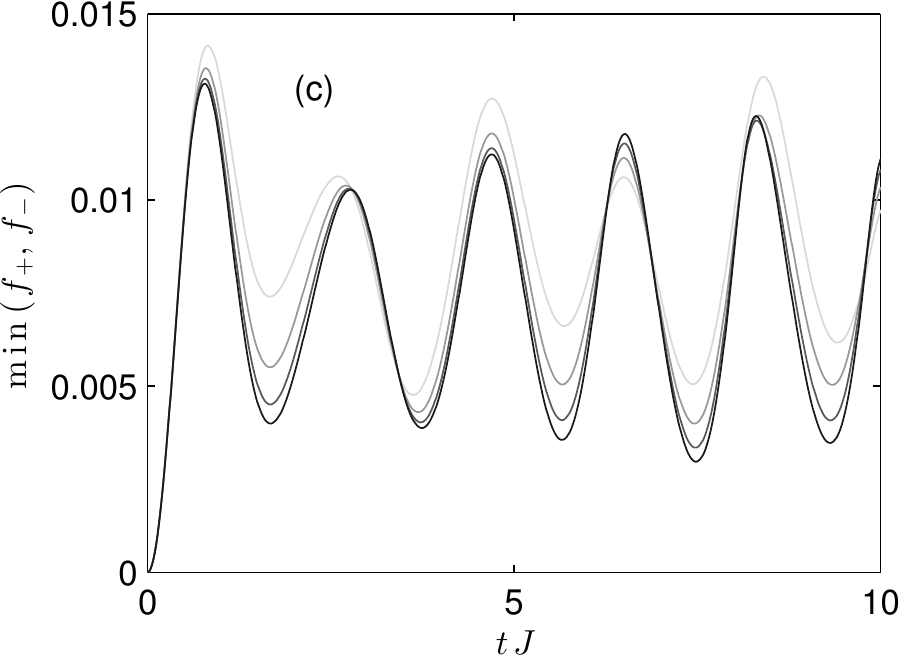}~~\includegraphics[width=0.4\textwidth]{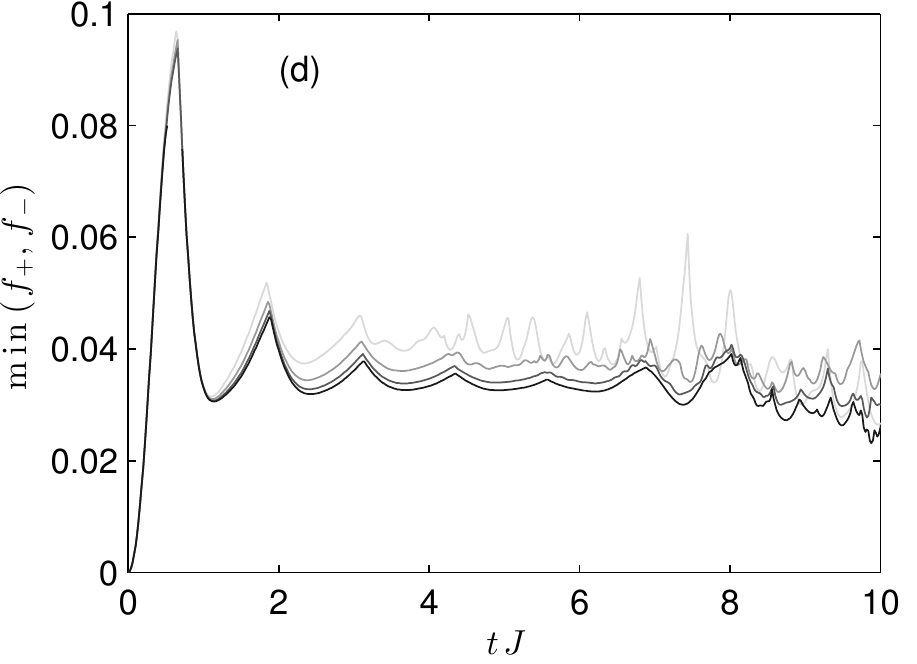}
\caption{Scaling of the return probability with the system size for quenches to $h_f=0.5$ (a,c) and $h_f=1.5$ (b,d). Panels (a,b) correspond to the interaction range $\alpha=1.8$, where we still see a non-vanishing time-averaged order parameter. The panels (c,d) show simulations for $\alpha=2.5,$ where the time-averaged order parameter vanishes, but the cusp singularities  in the return probability still persist. The colors correspond to different system sizes; $N=25, 50, 100, 200$ from bright to dark.} 
\label{fig10}
\end{figure}

In this section we study the scaling of the Loschmidt echo and the return probability with the system size. As for the dynamical order parameter, the understanding of the behaviour of these quantities developed in the infinite-range case turns out to give us clear guidelines to interpret the data obtain numerically for finite $\alpha$.  For small $\alpha$ singularities in the Loschmidt echo coincide with local minima of $\sx(t)$ and singularities in the return probability coincide with zeros of $\sx(t)$ (see \fref{fig11}). In contrast to the dynamical order parameter the singularities in the Loschmidt echo and the return probability persist for any value of $\alpha$ we have checked; see \fref{fig9} and \fref{fig10}. 

In summary, for any value of $\alpha$, the transition in the time-averaged order parameter, the cusps in the Loschmidt echo, and the cusps in the return probability indicate a change in symmetry of the trajectory in phase space of the total magnetisation of the system. For quenches below the critical point the magnetisation precesses to its final value along the longitudinal direction and for quenches above the critical point it precesses around the transverse direction. While the presence or absence of a DQPT-OP dependens strongly on $\alpha$ the transition due to a change in symmetry of the trajectories is robust and occurs also  for $\alpha>2$ as can be detected by observing the dynamics of $\sx$ (or also by the Loschmidt echo or the return probability). 

\newpage
\section{Methods}
In this section we provide a short presentation of the methods used to produce the numerical data, the detailed settings used for simulations and discuss the convergence with the bond dimension.
\subsection{Details of the simulation}
\begin{figure}[h!!]
\includegraphics[width=0.45\textwidth]{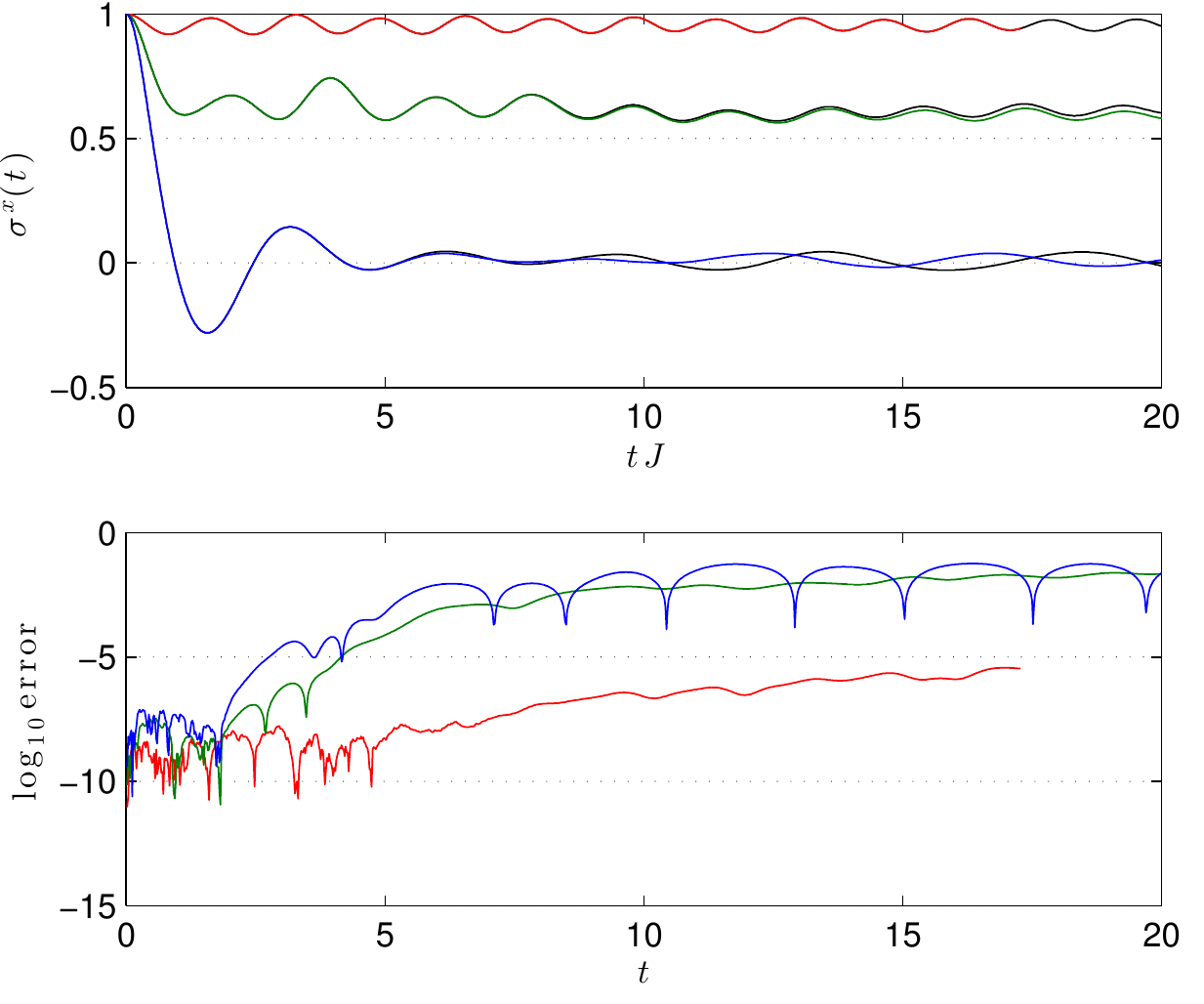}\includegraphics[width=0.45\textwidth]{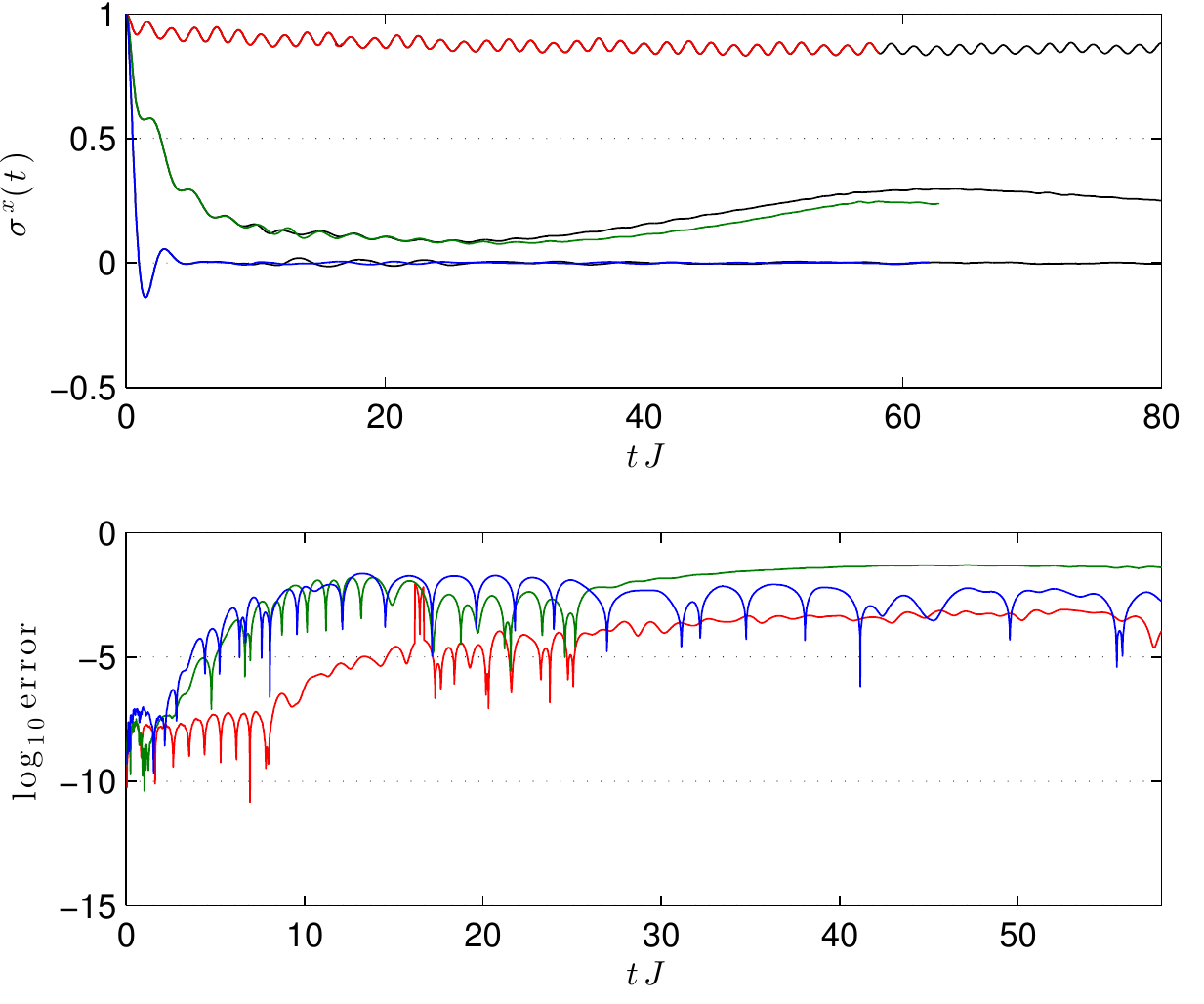}
\caption{Convergence of $\sx(t)$ for $\alpha=1.8$ (left) and $\alpha=2.5$ (right) with increasing  matrix product state bond dimension -- $D=60$ (black lines) and $D=100$ (colored lines). We show the data for $h_f$=0.4 (red), 0.8 (green), 1.2(blue). The system size is 100. The top plots show the time evolution of $\sx(t)$ for both bond dimensions and the bottom plot their difference, which is an estimate of the error as it is explained in \fref{fig7}. }
\label{fig6}
\end{figure}

\begin{figure}[h!!]
\begin{tabular}{c|cc}
&$h_f=0.5$&$h_f=1.5$\\
\hline
$\alpha=1.6$&\begin{minipage}{0.45\textwidth}\includegraphics[width=\textwidth]{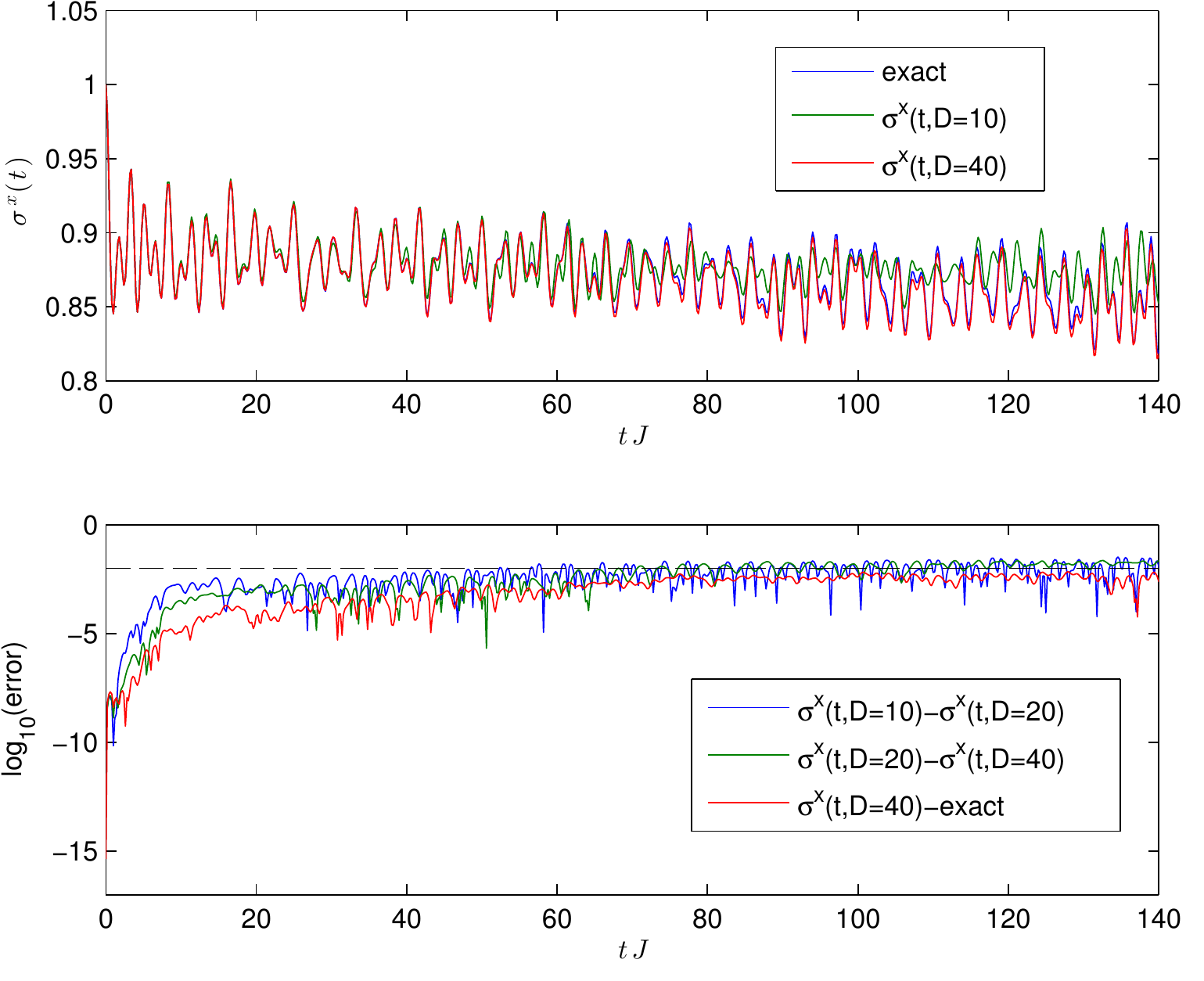}\end{minipage}&\begin{minipage}{0.45\textwidth}\includegraphics[width=\textwidth]{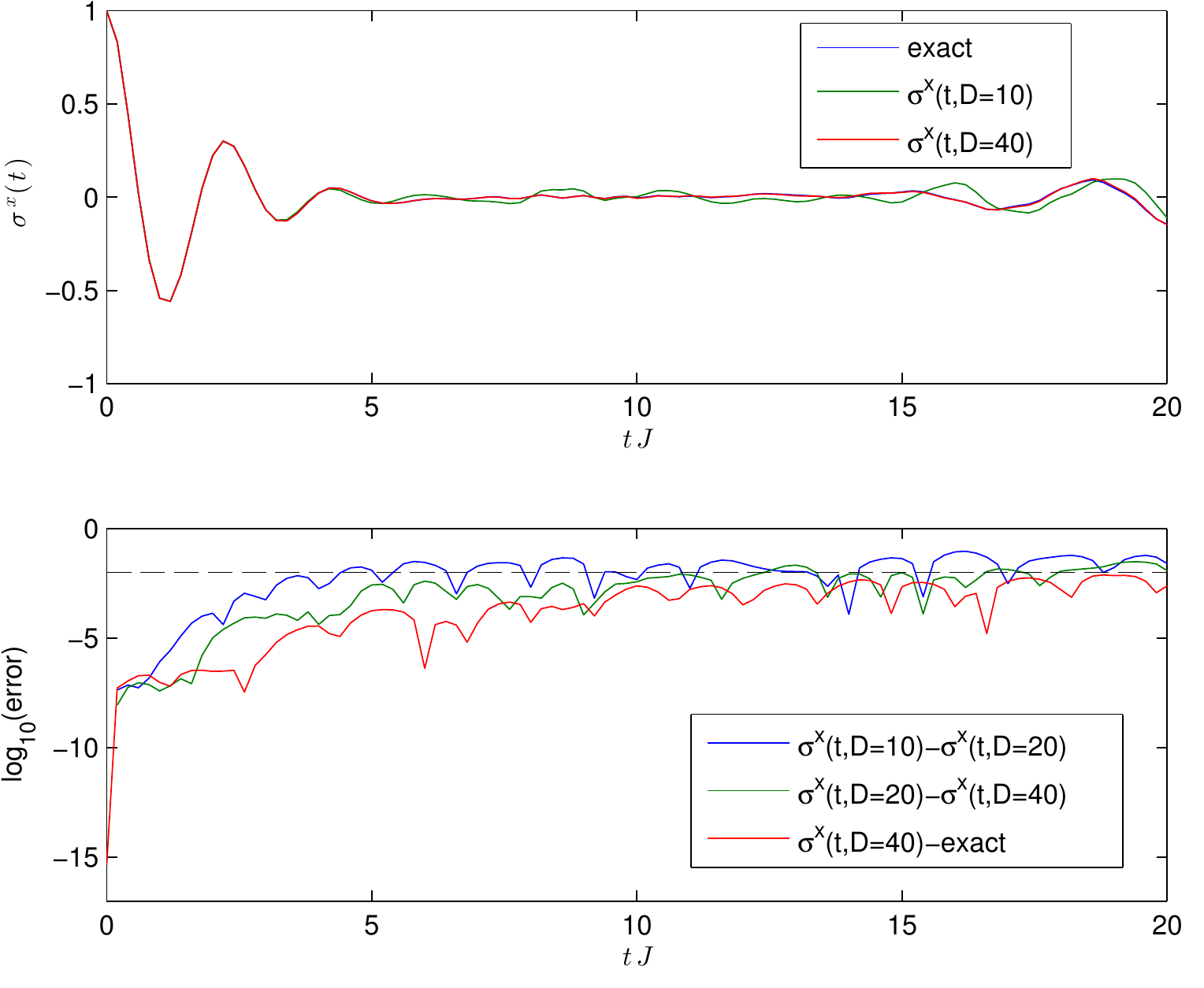}\end{minipage}\\
\hline
$\alpha=3.$&\begin{minipage}{0.45\textwidth}\includegraphics[width=\textwidth]{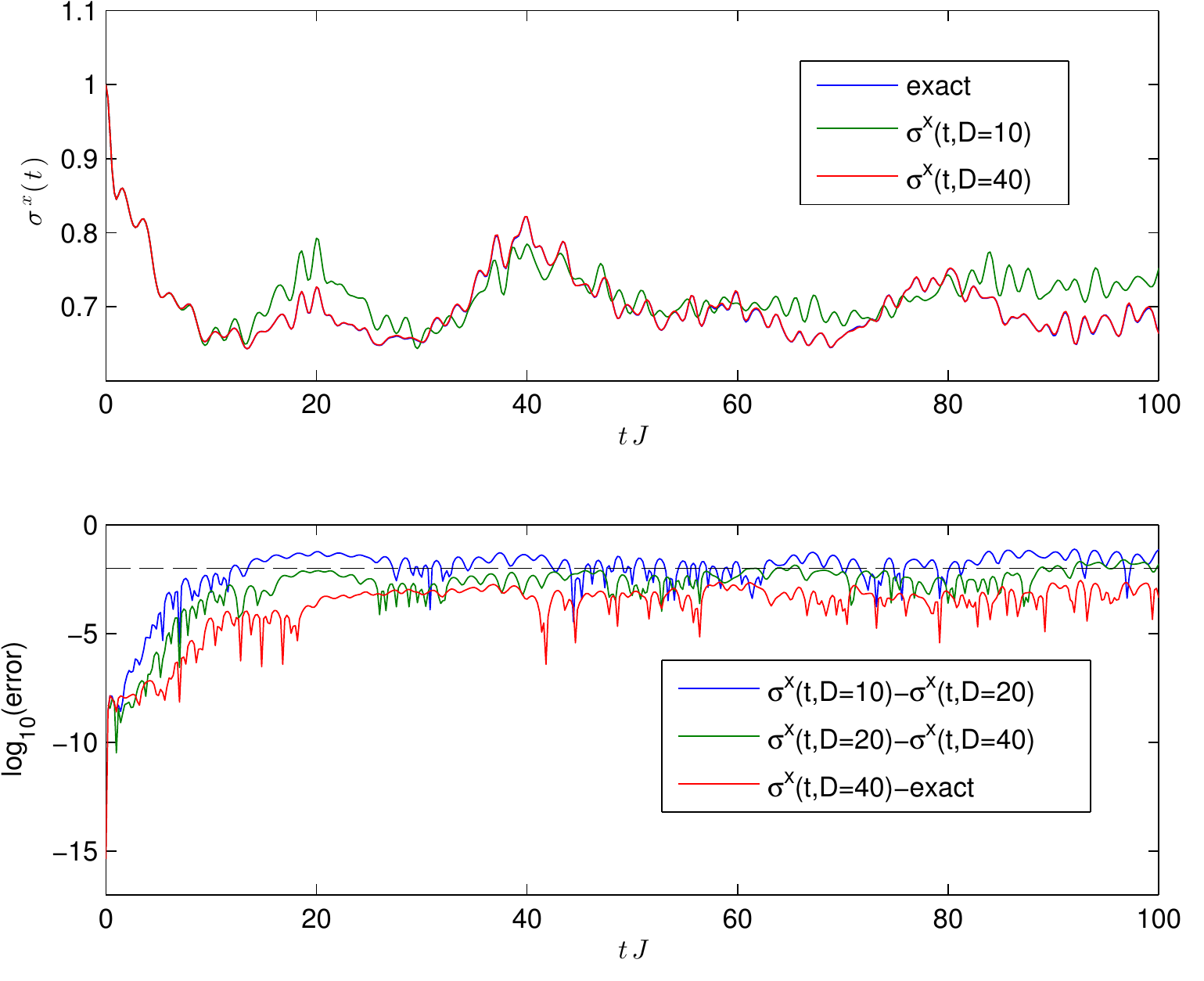}\end{minipage}&\begin{minipage}{0.45\textwidth}\includegraphics[width=\textwidth]{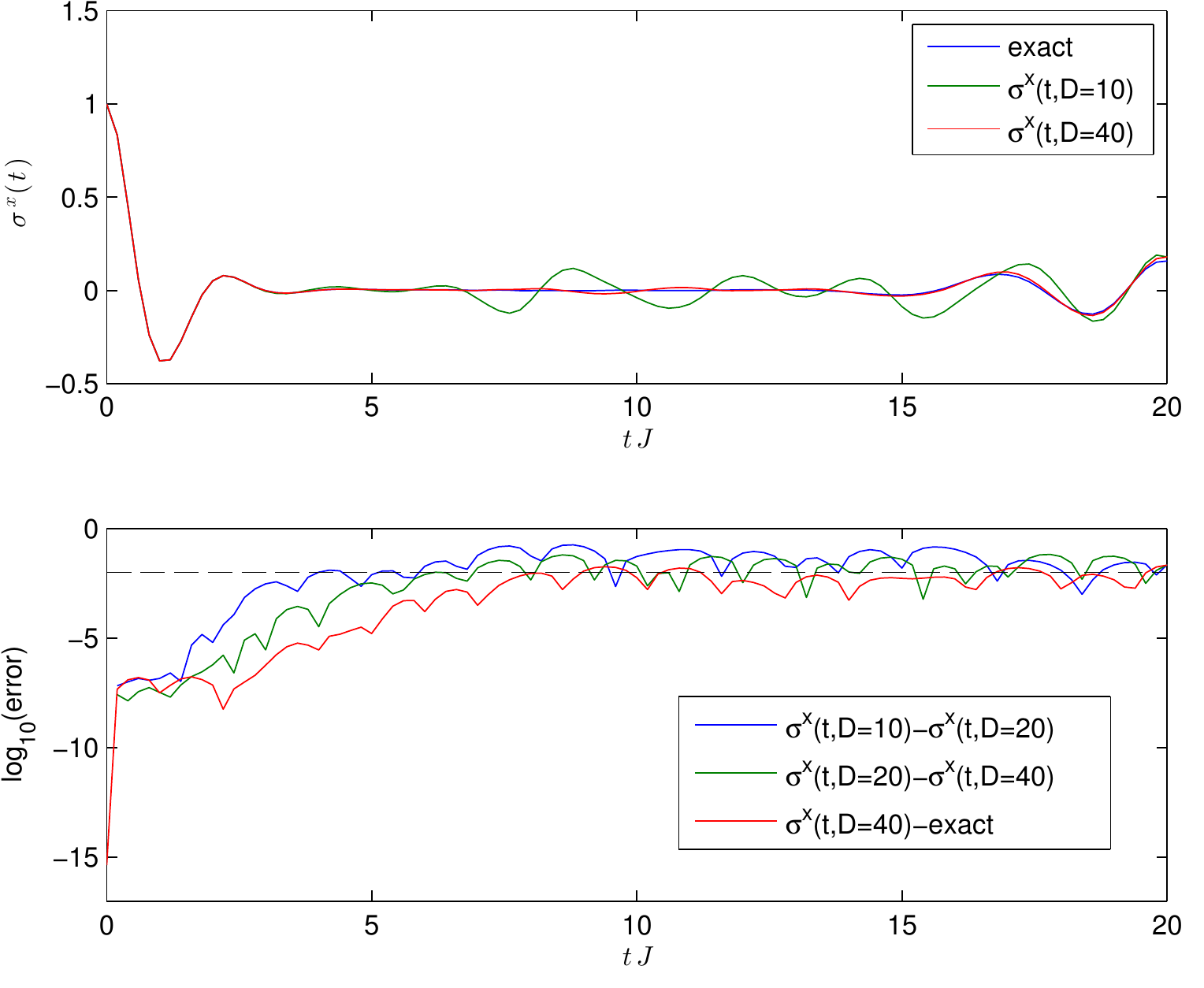}\end{minipage}
\end{tabular}
\caption{(Comparison of the TDVP calculations of the time evolution of the order parameter $\sx(t)$ with the exact results obtained by the Lanczos method.   In the rows and columns we change $\alpha$ and $h$ as indicated. We obtain an estimate of the error by calculating the difference of the expectations obtained for bond dimensions that differ by a factor of 2. We observe that the actual error is typically an order of magnitude smaller. }
\label{fig7}
\end{figure}

\begin{figure}[h!!]
\includegraphics[width=0.45\textwidth]{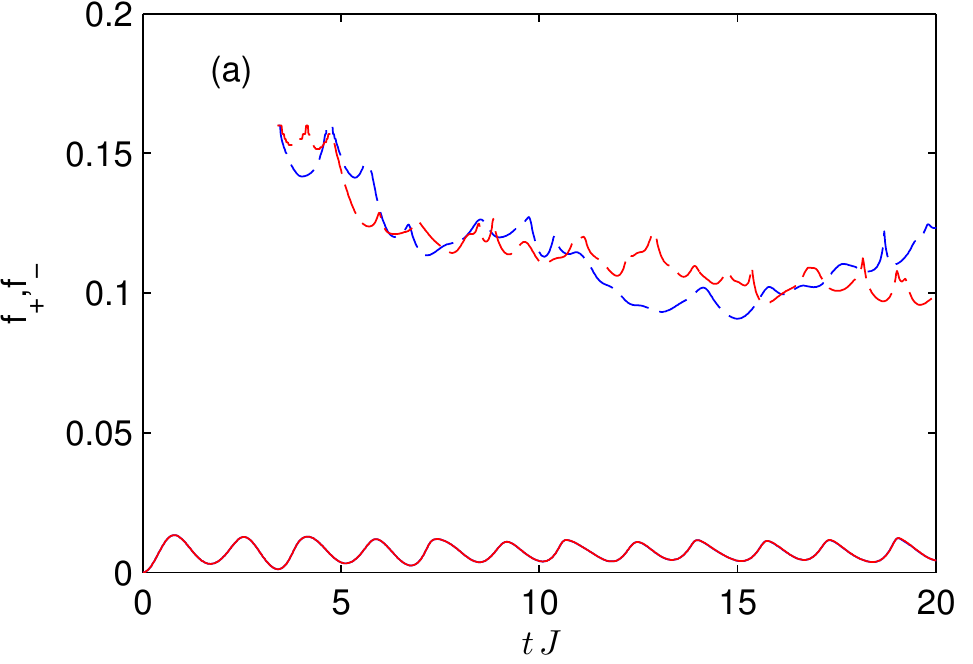}\includegraphics[width=0.45\textwidth]{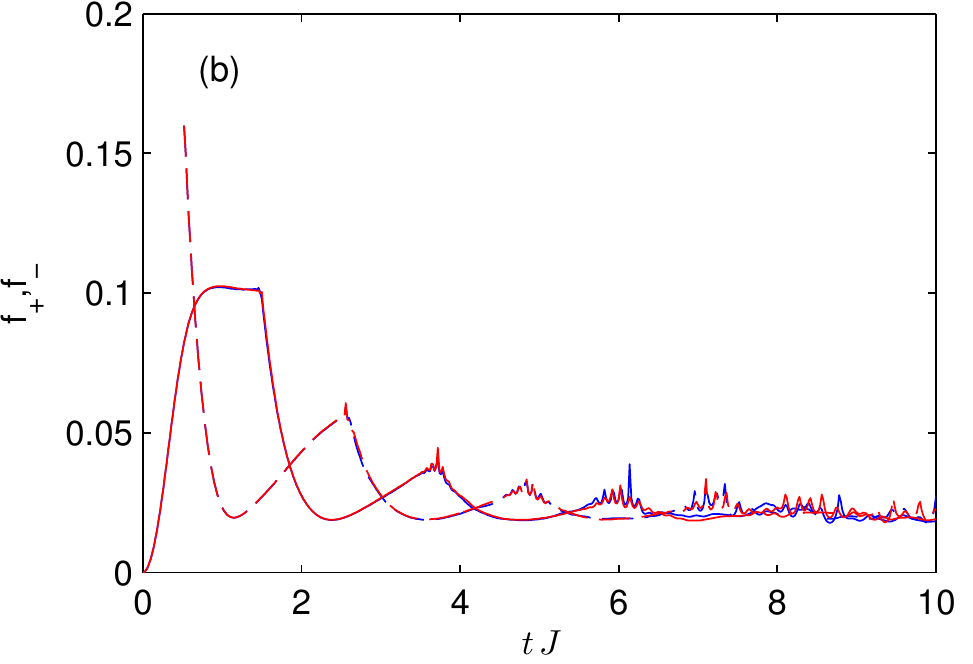}
\includegraphics[width=0.45\textwidth]{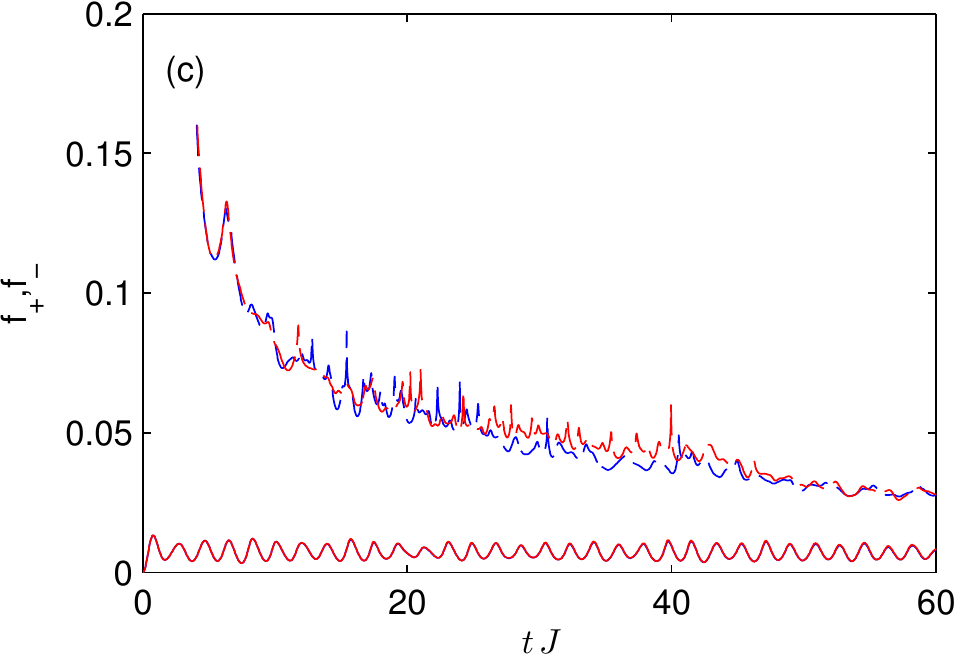}\includegraphics[width=0.45\textwidth]{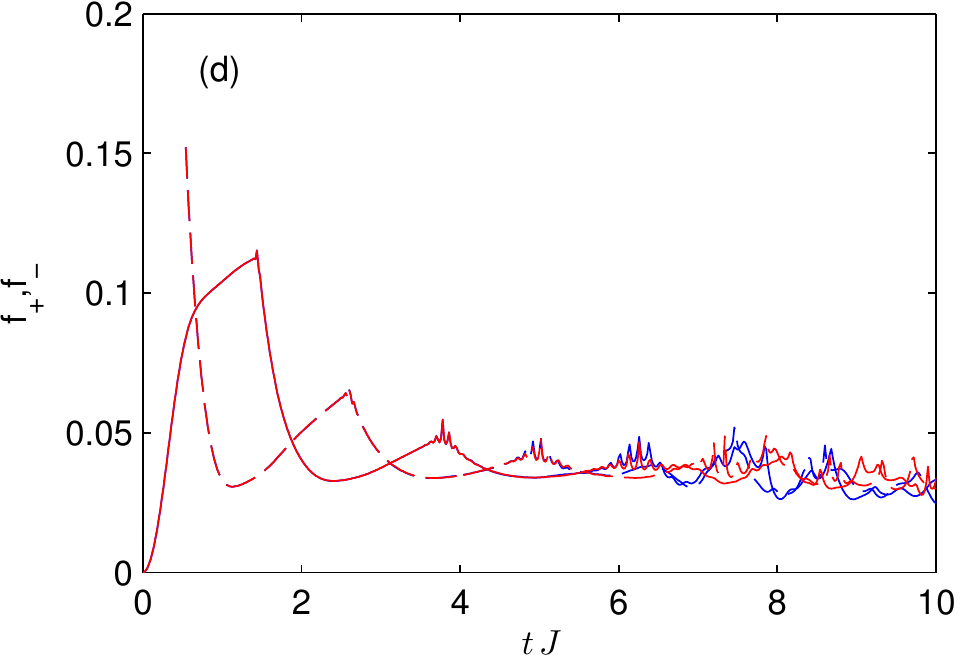}
\caption{Convergence of the probabilities to return to the symmetry-broken ground states (denoted by dashed and full lines) of the pre-quench Hamiltonian. We show data for $h=0.5$ (a,c) and $h=1.5$ (b,d) for interaction ranges $\alpha=1.8$ (a,b) and $\alpha=2.5$ (c,d). The color denotes different bond dimensions, namely $D=60$ (blue) and $D=100$ (red).}
\label{fig12}
\end{figure}

We use the variant of the matrix product state time dependent variational principle described in \cite{HLO+14} with the one-site second order integration method with step size 0.02. Further, we used a Hamiltonian matrix product operator description with a relative error smaller than $10^{-9}$. In order to check convergence of the data we used increasing bond dimension for the variational ansatz of the state. The largest bond dimension, for which the data in the main text is presented, is $D=100$. In \fref{fig6} we compare the expectation values of $\sx(t)$ for different $h_f$ and $\alpha=1.8, 2.5$ calculated for the bond dimensions $D=60,100$. The phase diagram shown in the main text is calculated from the data averaged over a window of $tJ=5$ around the "large" minimum before the revival (clearly seen for $\alpha=2.5$) for $h_f<1$ and around the latest converged window of size $tJ=5$.
In order to further justify our choice of times for which we compute the time-averaged order parameter we compare the results of our method to exact diagonalisation for small systems $N=18.$ In \fref{fig7} we show the time evolution of $\sx(t)$ along with the errors. We compare the estimated errors which are computed as a difference of the data calculated with bond dimensions that differ by a factor of two. We observe that the actual error is almost always around an order of magnitude smaller as the estimate. Besides, in the flat region where we extract the time average, the error does not grow significantly above $~0.01$, as the simulated curve does not start to deviate but rather oscillates around the exact value.
Finally, we check the convergence of the Loschmidt echo and the return probability rates. In \fref{fig12} we show the data for different quenches and ranges of interaction. For the considered time ranges the data converged with the bond dimension. 

\newpage
\subsection{Time dependent variational principle}
Recently, the time dependent density matrix renormalisation group (tDMRG) has been formulated in the language of the semiclassical phase space equations, by using the time dependent variational principle (TDVP) \cite{Haegeman2011}\cite{HLO+14}. Here we outline the basics of the TDVP.

The Schr\"odinger equation of motion can be obtained from the minimisation of the action
$$ \mathcal{S}=\int_{0}^tL(\bar{\psi},\psi,s)\dd s$$
with the Lagrangian (in the case of norm preserving states)
$$L=\frac{i}{2}\langle\psi|\dot{\psi} \rangle-\frac{i}{2}\langle\dot{\psi}|\psi \rangle-\langle \psi|H|\psi\rangle.$$
If we optimize the action $\mathcal{S}$ on the full Hilbert space $\mathcal{H}$ we obtain the Schr\"odinger equation. However, if we extremize the action on some restricted submanifold  $\mathcal{M}=\{\ket{\psi(z)},z\in\mathds{C}^n\}$ of $\mathcal{H}$, we arrive at Euler-Lagrange equations on the manifold which may be written as
$$\ii \dot{z_i}=G^{i\bar{j}}\langle \partial_{\bar{z}_j}\psi(\bar{z})|H|\psi(z)\rangle,$$
where (in the case of norm preserving states) $$G_{\bar{i},j}=\langle\partial_{\bar i}\psi|\partial_j\psi\rangle,\quad G^{i,\bar{j}}G_{\bar{j},k}=\delta^i_{k}.$$
If we further define the poisson bracket as 
$$\{f,g\}_{\rm P}=-\ii(\partial_{z_i}fG^{i,\bar{j}}\partial_{\bar{z}_j}g-\partial_{z_i}gG^{i,\bar{j}}\partial_{\bar{z}_j}f)$$
we can rewrite the evolution equation in the semiclassical form
$$\dot{z}_i=\{H,z_i\}_{\rm P}.$$
Application of the TDVP to the MPS manifold leads to a variant of DMRG with fixed bond dimension \cite{Haegeman2011,HLO+14}, which has a comparable efficiency to the standard formulation of tDMRG but in addition preserves all the symmetries of the Hamiltonian and is applicable to systems which admit a compact MPO description of the Hamiltonian. 
\newpage
\twocolumngrid
\bibliography{library}

\end{document}